\documentstyle[epsf]{mn}
\newif\ifAMStwofonts
\ifoldfss

  \ifCUPmtlplainloaded \else
    \NewTextAlphabet{textbfit} {cmbxti10} {}
    \NewTextAlphabet{textbfss} {cmssbx10} {}
    \NewMathAlphabet{mathbfit} {cmbxti10} {} % for math mode
    \NewMathAlphabet{mathbfss} {cmssbx10} {} %  "   "    "
  \fi
  \ifAMStwofonts
    \ifCUPmtlplainloaded \else
      \NewSymbolFont{upmath} {eurm10}
      \NewSymbolFont{AMSa} {msam10}
      \NewMathSymbol{\upi}     {0}{upmath}{19}
      \NewMathSymbol{\umu}     {0}{upmath}{16}
      \NewMathSymbol{\upartial}{0}{upmath}{40}
      \NewMathSymbol{\leqslant}{3}{AMSa}{36}
      \NewMathSymbol{\geqslant}{3}{AMSa}{3E}

       \let\le=\leqslant
       \let\ge=\geqslant
    \fi
  \fi
\fi % End of OFSS

\ifnfssone
  \newmathalphabet{\mathit}
  \addtoversion{normal}{\mathit}{cmr}{m}{it}
  \addtoversion{bold}{\mathit}{cmr}{bx}{it}
  \newmathalphabet{\mathbfit} % math mode version of \textbfit{..}
  \addtoversion{normal}{\mathbfit}{cmr}{bx}{it}
  \addtoversion{bold}{\mathbfit}{cmr}{bx}{it}
  \newmathalphabet{\mathbfss} % math mode version of \textbfss{..}
  \addtoversion{normal}{\mathbfss}{cmss}{bx}{n}
  \addtoversion{bold}{\mathbfss}{cmss}{bx}{n}
  \ifAMStwofonts
    \ifCUPmtlplainloaded \else
      \UseAMStwoboldmath
      \makeatletter
      \new@mathgroup\upmath@group
      \define@mathgroup\mv@normal\upmath@group{eur}{m}{n}
      \define@mathgroup\mv@bold\upmath@group{eur}{b}{n}
      \edef\UPM{\hexnumber\upmath@group}
      \new@mathgroup\amsa@group
      \define@mathgroup\mv@normal\amsa@group{msa}{m}{n}
      \define@mathgroup\mv@bold\amsa@group{msa}{m}{n}
      \edef\AMSa{\hexnumber\amsa@group}
      \makeatother
      \mathchardef\upi="0\UPM19
      \mathchardef\umu="0\UPM16
      \mathchardef\upartial="0\UPM40
      \mathchardef\leqslant="3\AMSa36
      \mathchardef\geqslant="3\AMSa3E

       \let\le=\leqslant
       \let\ge=\geqslant
    \fi
  \fi
\fi % End of NFSS release 1

\ifnfsstwo
  \DeclareMathAlphabet{\mathbfit}{OT1}{cmr}{bx}{it}
  \SetMathAlphabet\mathbfit{bold}{OT1}{cmr}{bx}{it}
  \DeclareMathAlphabet{\mathbfss}{OT1}{cmss}{bx}{n}
  \SetMathAlphabet\mathbfss{bold}{OT1}{cmss}{bx}{n}
  \ifAMStwofonts
    \ifCUPmtlplainloaded \else
      \DeclareSymbolFont{UPM}{U}{eur}{m}{n}
      \SetSymbolFont{UPM}{bold}{U}{eur}{b}{n}
      \DeclareSymbolFont{AMSa}{U}{msa}{m}{n}
      \DeclareMathSymbol{\upi}{0}{UPM}{"19}
      \DeclareMathSymbol{\umu}{0}{UPM}{"16}
      \DeclareMathSymbol{\upartial}{0}{UPM}{"40}
      \DeclareMathSymbol{\leqslant}{3}{AMSa}{"36}
      \DeclareMathSymbol{\geqslant}{3}{AMSa}{"3E}

       \let\le=\leqslant
       \let\ge=\geqslant
    \fi
  \fi
\fi % End of NFSS release 2
\ifCUPmtlplainloaded \else
  \ifAMStwofonts \else % If no AMS fonts
    \def\upi{\pi}
    \def\umu{\mu}
    \def\upartial{\partial}
  \fi
\fi

\title{Hidden Subluminous Stars among the FAUST  UV sources toward OPHIUCHUS}

\author [L. Formiggini et al.]
{Liliana Formiggini, Noah Brosch, Elchanan Almoznino
\\ The Wise Observatory and the School of Physics and Astronomy 
\\ Raymond and Beverly Sackler Faculty of Exact Sciences 
\\Tel Aviv University, Tel Aviv 69978, Israel;\\ lili@wise.tau.ac.il,
noah@wise.tau.ac.il, nan@wise.tau.ac.il 
\\ \and Stuart Bowyer, Michael Lampton 
\\ Space Sciences Laboratory  \\ University of California, 
\ Berkeley CA 94720-7450, U.S.A.; \\bowyer@mofo.ssl.berkley.edu, 
mlampton@ssl.berkley.edu\\}
\date{Accepted 2002 MMM DD,
 Received 2001 MMM DD,
      in original form 2001  MMM DD }

\pagerange{\pageref{firstpage}--\pageref{lastpage}}
\pubyear{2002}

\begin{document}

\def\etal{{\it et al.\ }}
\def\kms{$\rm km\, s^{-1}$}
\def\msol{M$_{\odot}$ }
\def\eg{{\it e.g.,}}
\def\ie{{\it i.e.,}}
\def\Halpha{H$\alpha$}
\def\Hbeta{H$\beta$}
\def\Hgamma{H$\gamma$}
\def\Hdelta{H$\delta$}
\def\Lya{Ly$\alpha$}
\def\Lyb{Ly$\beta$}

\maketitle

\label{firstpage}

\begin{abstract}
We present results of an analysis of a UV image in the direction of  Ophiuchus,
obtained with  the FAUST instrument. The image contains 228 UV sources. Most 
of these are identified as normal early-type stars through correlations 
with cataloged objects. For the first time in this project we identify
UV sources as such stars by selecting suitable candidates  in crowded fields as 
the bluest objects in color-color diagrams using observations from the Wise
Observatory. These candidates are then studied using  
low-resolution spectroscopy, which allowed the determination
of spectral types to an accuracy of about one-half class, for 60 stars.

Synthetic photometry of spectral data is performed in order to predict 
the expected UV emission, on the basis of the photometric information.
These results are used along with the Hipparcos/Tycho 
information, to search for sub-luminous stars.
The comparison of the predicted emission with the FAUST 
measured magnitudes allows us to select twelve stars as  
highly-probable evolved hot stars.
High signal-to-noise spectra are obtained for nine of these stars and 
Balmer line profiles were compared with the prediction of atmosphere 
models and with the spectrum of  real stellar atmospheres.
Among the nine candidates,  six are classified as previously 
unrecognized sdB stars and two as white dwarfs.
Our result indicates that indeed more bright subluminous stars are 
still unrecognized in the existing samples.
\end{abstract}


\section{Introduction}

In the vacuum ultraviolet range observations are very scarce
and are limited to very bright sources, such as those measured by  the
TD-1 mission (Thompson et al. 1978).
The FAUST experiment (Bowyer et al. 1993)  represents a successful attempt 
to produce
a catalog of fainter UV sources in  a few regions of the sky.  
During the FAUST mission, about 5\% of the celestial sphere was observed.
First results were published in the FAUST source catalog (FSC: Bowyer et al. 1995).

One of the fields
observed during the FAUST mission on board  the ATLAS-1 Shuttle
mission (Lampton et al. 1993) is located in the outer region of the
Ophiuchus molecular cloud.  Among the fields observed by FAUST,
this is the UV field nearest the galactic equator.

The Ophiuchus molecular cloud region
%, centered at 16$^h$25$^m$,-24$^o$00',
subtends several tens of square degrees on the sky, at a distance 
of 125 pc. The region observed most intensively within this complex is the
$\rho$ Ophiuchus core, which is a star formation region.
 In addition to the core, there are 
diffuse and filamentary dust structures, as seen on the Palomar Sky Survey(PSS) plates.
Owing to the location  near the galactic 
center direction (l $\sim $355 and b $\sim $15), with a large number of
projected stars, this is a very crowded region.

This paper presents a list of the sources detected in this
field and their optical identification, obtained by correlations with 
published catalogues (section 2 and 3) and by photometric and spectroscopic
observations performed at the Wise Observatory (section 4). 
The results are used below, along with Hipparcos/Tycho results, to 
identify new hot, sub-luminous stars (section 5).
Synthetic photometry using  spectral data in the ultraviolet domain 
allows us to predict the expected UV brightness of a source. 
Comparing this to the detected UV brightness yields  a list of ten
objects much brighter in the UV than expected. Nine of these candidates 
were followed up with spectroscopy to study the Balmer line profiles.
We propose that six of these are hot subdwarfs and two are possibly 
white dwarfs, which were previously misidentified as normal 
main-sequence stars.

\section{FAUST observations of Ophiuchus}

FAUST (FAr Ultraviolet Space Telescope) is a wide-field ($\sim $ 8$^0$
field of view) telescope with a bandpass from 1400 \AA\  to 1800 \AA\      
and an angular resolution  of 3.5 arc minutes, designed for imaging
in the far ultraviolet.  The FAUST telescope operated on board of the
space shuttle  Atlantis during the ATLAS-1 mission and imaged 22 sky
fields.
Details of the image reconstruction and  performance are given in Bowyer et al. (1993).
A catalog of the 4698 UV sources detected by FAUST and of their optical
identification, based on positional coincidence with entries from catalogues of
stars and galaxies, was produced by Bowyer et al. (1995). Most of the 22 
fields were analyzed at the Tel Aviv  University and accurate identifications
were  obtained with the help of ground-based spectroscopy.
The results of this systematic analysis have already been presented for the following
sky regions: the North Galactic Pole (Brosch et al. 1995); the Virgo cluster
region (Brosch et al. 1997); the Coma Cluster region (Brosch et al. 1998); 
the Fourth Galactic Quadrant region (Brosch et al. 2000a); a region near the  
North Galactic Pole (Brosch et al. 2000b) and the Antennae and NGC 6752 regions
 (Daniels et al. 2001). 
One of the fields observed by FAUST is  centered at 
$\alpha$$_{1950}$=17$^h$20$^m$, $\delta$$_{1950}$=-20$^0$00', on the outskirts 
of the Ophiuchus cloud, 14$^0$.3 NE from the center of the molecular cloud,
in a very crowded, low galactic latitude region.

The total sky area covered by the FAUST Ophiuchus image 
is $\sim $ 49.4 square degrees ($\Box^{\circ}$).
We analyzed this image using an impartial  source detector algorithm  
(Brosch et al. 1995), that was  adopted for the  FAUST fields already 
analyzed at Tel-Aviv.
This algorithm takes into account the different image depths that affect
various regions of the image and is characterized by a signal-to-noise 
threshold (Brosch et al. 1997). 
This impartial automatic algorithm detected 228 UV sources in the Ophiuchus
FAUST frame. The FAUST image is shown in Fig 1.
\begin{figure}
%\vspace{12cm}
 \centerline{\epsfxsize=3.0in\epsfbox{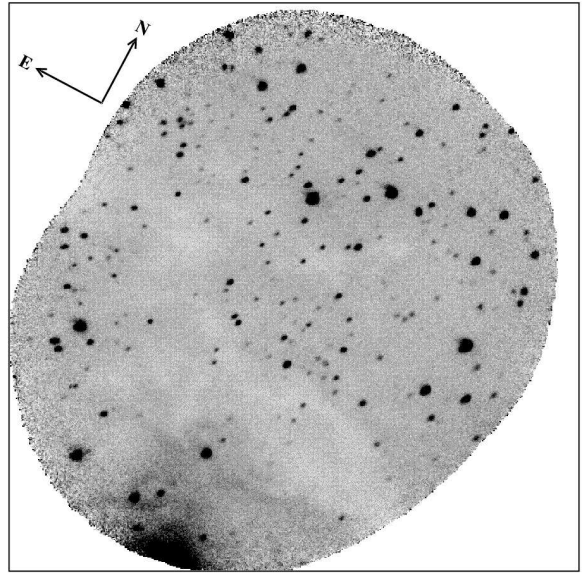}}
\caption{The FAUST image of the Ophiuchus field. The elongated stellar images are the 
results of imperfect alignment of the time-tagged positions. Note the lightly
colored patches near the center and in the Southern part of the image; these are
dust clouds blocking part of the background.}
\end{figure}

Table 1 lists the coordinates of the detected sources 
with their calculated coordinates and their UV magnitude. The  UV 
monochromatic magnitude  is defined as  
\begin{equation}
m_{UV}=-21.175-2.5*log(FD)
\end{equation}
where the flux density FD is in erg/sec/cm$^2$/\AA. The flux density was
derived from the count rate of FAUST, as explained in Brosch et al. (1995).
The error in the flux takes into account both the instrumental error and 
the systematic one resulting from the laboratory calibration of FAUST,
as explained in Brosch et al. (1997).
Fig 2 shows the distribution in UV flux  of the detected sources expressed
as monochromatic magnitudes.
\begin{figure}
\hspace{10cm}
\centerline{\epsfxsize=3.5in\epsfbox{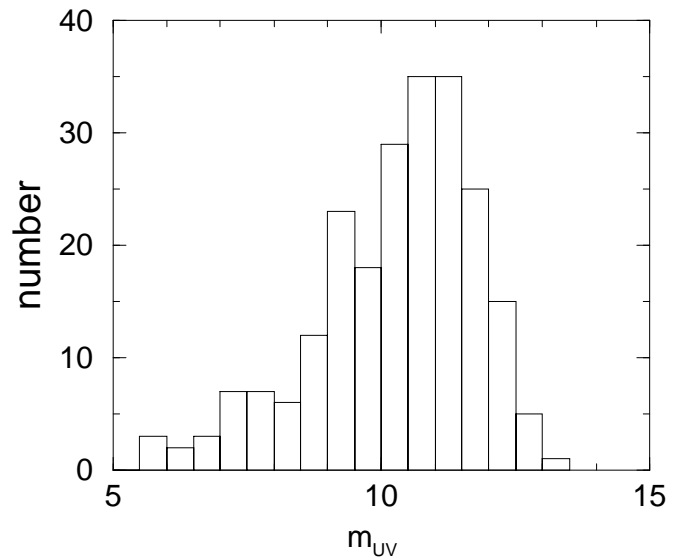}}
\caption{Histogram of the number of sources versus the measured Faust magnitudes.}
\end{figure}
As previously mentioned, FAUST  sources in this  field have already been
extracted by Bowyer et al. (1995)-FSC using  a  flux-limited algorithm.
We compared our detections with that from FSC and found that all 
but nine sources have already been  listed in the FSC. Although the detection methods
are different, we  believe that these additional sources are true detections.  
The last column of Table 1 lists the number of the source in the 
FSC catalog. An asterisk marks the nine new detections.

\begin{table*}
 \caption{Detected FAUST sources in the Ophiuchus field}
%\begin{tabular}{ll}
%\\
\begin{tabular}{@{}rrrrc@{}}
Id.no.& $\alpha$$_{2000}$ &$\delta$$_{2000}$ & m$_{UV}$&FAUST no. \\
\\
   1&  17 34 52.46&  -16 38 47.44&   7.66$\pm$ 0.20&  4219\\ 
   2&  17 30 02.86&  -16 03 02.95&   8.86$\pm$ 0.20&  4174\\ 
   3&  17 31 05.91&  -16 11 30.41&  11.03$\pm$ 0.44&  4183\\ 
   4&  17 29 06.64&  -16 02 38.11&  10.92$\pm$ 0.36&  4167\\ 
   5&  17 32 37.55&  -16 39 07.30&   9.01$\pm$ 0.20&  4196\\ 
   6&  17 29 24.29&  -16 36 25.46&   7.42$\pm$ 0.17&  4169\\ 
   7&  17 34 12.77&  -17 09 18.44&   9.15$\pm$ 0.20&  4211\\ 
   8&  17 33 44.52&  -17 07 13.78&   9.58$\pm$ 0.21&  4204\\ 
   9&  17 23 45.05&  -16 03 46.08&  10.90$\pm$ 0.29&  4127\\ 
  10&  17 31 17.31&  -17 09 23.53&   7.06$\pm$ 0.17&  4184\\ 
  11&  17 37 44.70&  -17 51 27.32&   7.58$\pm$ 0.19&  4242\\ 
  12&  17 25 38.48&  -16 28 00.63&  11.89$\pm$ 0.47&  4142\\ 
  13&  17 25 38.87&  -16 36 50.61&  11.57$\pm$ 0.42&  4143\\ 
  14&  17 35 41.52&  -17 50 09.55&  11.07$\pm$ 0.36&  4227\\ 
  15&  17 23 23.91&  -16 29 18.81&  11.02$\pm$ 0.31&  4126\\ 
  16&  17 18 38.14&  -16 01 25.40&   8.80$\pm$ 0.19&  4077\\ 
  17&  17 39 18.40&  -18 25 47.55&   8.52$\pm$ 0.23&  4250\\ 
  18&  17 34 04.30&  -17 51 11.68&  11.54$\pm$ 0.34&  4206\\ 
  19&  17 28 42.29&  -17 15 39.92&   8.63$\pm$ 0.17&  4162\\ 
  20&  17 30 33.18&  -17 29 36.13&  11.24$\pm$ 0.35&  4179\\ 
  21&  17 35 15.05&  -18 03 16.67&  10.86$\pm$ 0.31&  4221\\ 
  22&  17 26 52.57&  -17 06 55.83&  10.85$\pm$ 0.28&  4155\\ 
  23&  17 34 49.35&  -18 05 42.44&  10.99$\pm$ 0.26&  4218\\ 
  24&  17 32 18.21&  -17 48 53.36&  12.28$\pm$ 0.49&  4193\\ 
  25&  17 28 54.02&  -17 23 36.28&   9.17$\pm$ 0.18&  4164\\ 
  26&  17 29 47.23&  -17 36 13.25&   9.86$\pm$ 0.20&  4171\\ 
  27&  17 35 22.87&  -18 15 56.02&   9.17$\pm$ 0.18&  4224\\ 
  28&  17 25 28.88&  -17 05 59.25&  11.44$\pm$ 0.35&  4140\\ 
  29&  17 36 20.24&  -18 25 13.77&   9.79$\pm$ 0.22&  4233\\ 
  30&  17 39 06.33&  -18 44 00.14&   9.34$\pm$ 0.22&  4249\\ 
  31&  17 33 23.46&  -18 03 09.05&  12.10$\pm$ 0.50&  4200\\ 
  32&  17 34 40.00&  -18 14 22.22&   9.96$\pm$ 0.21&  4215\\ 
  33&  17 21 08.67&  -16 38 48.83&  10.79$\pm$ 0.27&  4103\\ 
  34&  17 16 14.01&  -16 04 26.31&   9.74$\pm$ 0.21&  4057\\ 
  35&  17 35 06.68&  -18 19 40.59&   9.14$\pm$ 0.18&  4220\\ 
  36&  17 32 13.03&  -18 02 37.64&  12.05$\pm$ 0.47&  4191\\ 
  37&  17 20 03.40&  -16 40 51.02&   8.35$\pm$ 0.17&  4092\\ 
  38&  17 19 35.72&  -16 32 49.49&  11.68$\pm$ 0.41&  *\\    
  39&  17 35 57.63&  -18 35 07.11&  10.08$\pm$ 0.22&  4228\\ 
  40&  17 14 28.13&  -15 57 01.76&   9.67$\pm$ 0.20&  4042\\ 
  41&  17 16 08.68&  -16 14 02.60&   9.57$\pm$ 0.22&  4056\\ 
  42&  17 30 10.18&  -17 56 16.81&  11.31$\pm$ 0.28&  4175\\ 
  43&  17 39 32.25&  -19 01 26.78&  10.32$\pm$ 0.33&  4252\\ 
  44&  17 33 30.15&  -18 20 21.28&  12.03$\pm$ 0.43&  4201\\ 
  45&  17 26 33.76&  -17 36 09.30&  11.53$\pm$ 0.32&  4149\\ 
  46&  17 32 23.70&  -18 14 41.78&  12.63$\pm$ 0.69&  *\\    
  47&  17 34 24.73&  -18 37 05.44&   9.64$\pm$ 0.19&  4214\\ 
  48&  17 18 14.01&  -16 40 22.15&  10.87$\pm$ 0.28&  4071\\ 
  49&  17 33 25.99&  -18 31 32.59&  11.04$\pm$ 0.26&  4202\\ 
  50&  17 38 27.88&  -19 05 14.45&  10.99$\pm$ 0.34&  4245\\ 
  51&  17 22 03.21&  -17 13 59.85&   9.00$\pm$ 0.18&  4111\\ 
  52&  17 15 35.01&  -16 26 17.91&  10.56$\pm$ 0.29&  4051\\ 
  53&  17 22 24.52&  -17 20 46.12&   7.57$\pm$ 0.17&  4114\\ 
  54&  17 29 46.64&  -18 08 30.52&  11.94$\pm$ 0.37&  4172\\ 
  55&  17 34 20.36&  -18 47 01.07&   9.03$\pm$ 0.18&  4212\\ 
  56&  17 28 14.08&  -18 00 41.09&  13.13$\pm$ 0.74&  4161\\ 
  57&  17 26 46.17&  -17 53 13.72&  10.64$\pm$ 0.23&  4153\\ 
  58&  17 20 24.82&  -17 12 56.01&  10.24$\pm$ 0.22&  4094\\ 
  59&  17 20 53.96&  -17 19 10.32&  11.47$\pm$ 0.34&  *\\    
  \end{tabular}

\end{table*}
\begin{table*}

%&
\begin{tabular}{@{}rrrrc@{}}
 Id no. & $\alpha$$_{2000}$ &$\delta$$_{2000}$& m$_{UV}$ &FAUST no. \\
\\
  60&  17 26 32.59&  -18 05 38.41&  10.50$\pm$ 0.23&  4150\\ 
  61&  17 37 02.98&  -19 21 17.12&  10.70$\pm$ 0.27&  4237\\ 
  62&  17 32 33.13&  -18 50 57.63&  12.62$\pm$ 0.71&  4195\\ 
  63&  17 22 35.72&  -17 42 18.03&   9.56$\pm$ 0.19&  4118\\ 
  64&  17 12 52.33&  -16 29 27.72&  11.69$\pm$ 0.49&  4036\\ 
  65&  17 29 43.35&  -18 33 53.18&  10.32$\pm$ 0.21&  4273\\ 
  66&  17 26 42.46&  -18 15 53.31&  10.56$\pm$ 0.22&  4151\\ 
  67&  17 31 36.29&  -18 51 01.23&  13.75$\pm$ 1.16&  4188\\ 
  68&  17 29 26.98&  -18 40 30.56&   8.73$\pm$ 0.17&  4168\\ 
  69&  17 23 25.09&  -17 58 09.90&   9.64$\pm$ 0.20&  4125\\ 
  70&  17 11 26.94&  -16 28 30.50&   9.44$\pm$ 0.22&  4027\\ 
  71&  17 25 19.95&  -18 16 56.65&   8.35$\pm$ 0.17&  4138\\ 
  72&  17 32 21.36&  -19 03 37.49&  11.61$\pm$ 0.34&  4192\\ 
  73&  17 22 19.42&  -17 53 10.31&  12.18$\pm$ 0.54&  4113\\ 
  74&  17 28 08.31&  -18 36 38.64&  15.55$\pm$ 6.75&  4160\\ 
  75&  17 19 54.48&  -17 46 14.21&   5.64$\pm$ 0.16&  4088\\ 
  76&  17 33 15.22&  -19 22 48.21&   9.68$\pm$ 0.20&  4199\\ 
  77&  17 16 08.51&  -17 18 32.16&  10.74$\pm$ 0.27&  4055\\ 
  78&  17 21 15.58&  -18 02 45.70&   9.08$\pm$ 0.18&  4104\\ 
  79&  17 24 37.01&  -18 27 32.35&   5.63$\pm$ 0.16&  4133\\ 
  80&  17 22 50.54&  -18 11 32.59&  12.02$\pm$ 0.46&  4120\\ 
  81&  17 37 35.10&  -20 03 59.62&  10.47$\pm$ 0.23&  4240\\ 
  82&  17 17 02.15&  -17 38 52.84&   9.42$\pm$ 0.18&  4063\\ 
  83&  17 35 36.04&  -19 54 08.54&   9.42$\pm$ 0.18&  4225\\ 
  84&  17 23 27.53&  -18 26 10.14&  11.14$\pm$ 0.32&  4124\\ 
  85&  17 17 35.65&  -17 50 52.15&   7.76$\pm$ 0.18&  4066\\ 
  86&  17 14 19.60&  -17 27 02.94&   7.21$\pm$ 0.17&  4040\\ 
  87&  17 12 12.28&  -17 13 37.90&   7.48$\pm$ 0.18&  4030\\ 
  88&  17 27 01.10&  -18 59 20.70&  10.45$\pm$ 0.21&  4156\\ 
  89&  17 38 35.48&  -20 19 01.36&  11.59$\pm$ 0.42&  4246\\ 
  90&  17 24 26.69&  -18 50 18.71&   9.22$\pm$ 0.18&  4130\\ 
  91&  17 30 37.91&  -19 33 34.58&  10.77$\pm$ 0.26&  4178\\ 
  92&  17 16 48.29&  -17 55 45.54&  11.54$\pm$ 0.41&  4061\\ 
  93&  17 27 49.01&  -19 16 47.86&  11.30$\pm$ 0.31&  4159\\ 
  94&  17 39 21.47&  -20 44 02.54&   8.92$\pm$ 0.16&  4251\\ 
  95&  17 34 24.41&  -20 06 24.04&  11.04$\pm$ 0.26&  4213\\ 
  96&  17 15 31.49&  -17 50 31.23&  11.68$\pm$ 0.43&  4049\\ 
  97&  17 19 43.43&  -18 21 47.62&  12.04$\pm$ 0.59&  4086\\ 
  98&  17 10 17.38&  -17 19 08.68&   9.97$\pm$ 0.23&  4021\\ 
  99&  17 12 27.86&  -17 35 08.11&  10.71$\pm$ 0.27&  4032\\ 
 100&  17 25 45.63&  -19 14 27.95&  10.78$\pm$ 0.25&  4144\\ 
 101&  17 37 55.40&  -20 40 58.58&   9.01$\pm$ 0.18&  4243\\ 
 102&  17 19 13.73&  -18 31 06.46&  11.76$\pm$ 0.46&  4082\\ 
 103&  17 35 50.46&  -20 29 43.22&  11.96$\pm$ 0.49&  4229\\ 
 104&  17 15 42.83&  -18 10 22.73&  11.32$\pm$ 0.34&  4052\\ 
 105&  17 20 15.65&  -18 49 24.94&   8.07$\pm$ 0.18&  4093\\ 
 106&  17 26 21.82&  -19 30 34.55&  10.34$\pm$ 0.21&  4147\\ 
 107&  17 38 51.22&  -20 58 58.05&   9.22$\pm$ 0.20&  4247\\ 
 108&  17 20 52.91&  -18 54 05.27&   9.67$\pm$ 0.19&  4100\\ 
 109&  17 22 28.75&  -19 05 54.60&  10.12$\pm$ 0.23&  4115\\ 
 110&  17 35 26.35&  -20 39 48.98&  10.30$\pm$ 0.23&  4223\\ 
 111&  17 14 27.28&  -18 11 25.52&  11.47$\pm$ 0.33&  4041\\ 
 112&  17 09 02.26&  -17 34 15.83&   8.03$\pm$ 0.19&  4014\\ 
 113&  17 34 14.75&  -20 33 28.26&  11.79$\pm$ 0.43&  4209\\ 
 114&  17 13 53.10&  -18 12 12.66&  10.80$\pm$ 0.25&  4039\\ 
 115&  17 36 51.12&  -20 57 45.45&  11.21$\pm$ 0.31&  4236\\ 
 116&  17 35 41.66&  -20 49 42.58&  11.42$\pm$ 0.29&  4226\\ 
 117&  17 15 25.96&  -18 23 44.05&  12.02$\pm$ 0.48&   *\\   
 118&  17 12 27.46&  -18 07 32.86&   7.69$\pm$ 0.18&  4033\\ 
\\
  \end{tabular}
\end{table*}
%\end{tabular}

%\clearpage
%\newpage
%\contcaption{}
%\begin{tabular}{ll}
%\\
\begin{table*}

\begin{tabular}{@{}rrrrc@{}}
Id.no. & $\alpha$$_{2000}$ &$\delta$$_{2000}$ & m$_{UV}$ & FAUSTno. \\
\\ 
 119&  17 15 00.35&  -18 24 12.47&  12.50$\pm$ 0.66&  4045\\ 
 120&  17 23 15.15&  -19 26 48.78&  10.40$\pm$ 0.23&  4123\\ 
 121&  17 40 17.67&  -21 21 51.19&  12.88$\pm$ 1.51&  *\\    
 122&  17 25 26.73&  -19 43 55.33&  10.71$\pm$ 0.23&  4139\\ 
 123&  17 37 40.33&  -21 11 24.90&  10.20$\pm$ 0.22&  4241\\ 
 124&  17 26 45.91&  -19 58 41.64&  12.61$\pm$ 0.55&  4152\\ 
 125&  17 11 04.39&  -18 09 26.68&  10.80$\pm$ 0.27&  4024\\ 
 126&  17 34 46.89&  -21 03 16.45&  10.24$\pm$ 0.23&  4216\\ 
 127&  17 27 11.82&  -20 18 10.39&   8.86$\pm$ 0.17&  4157\\ 
 128&  17 24 33.17&  -19 56 18.99&  11.84$\pm$ 0.45&  4131\\ 
 129&  17 17 31.16&  -19 08 33.83&  11.35$\pm$ 0.31&  4065\\ 
 130&  17 12 40.39&  -18 32 32.76&  11.63$\pm$ 0.36&  4035\\ 
 131&  17 11 57.93&  -18 27 37.48&  11.47$\pm$ 0.37&  4028\\ 
 132&  17 37 27.37&  -21 33 19.55&   9.02$\pm$ 0.19&  4238\\ 
 133&  17 08 29.88&  -18 11 35.95&   7.85$\pm$ 0.18&  4011\\ 
 134&  17 36 56.88&  -21 33 13.16&   9.50$\pm$ 0.19&  4235\\ 
 135&  17 36 24.99&  -21 29 07.67&  10.86$\pm$ 0.27&  4232\\ 
 136&  17 27 12.83&  -20 27 02.22&  10.32$\pm$ 0.22&  4158\\ 
 137&  17 09 08.91&  -18 16 33.79&  10.19$\pm$ 0.25&  4011\\ 
 138&  17 29 01.18&  -20 44 16.02&  12.13$\pm$ 0.40&  4166\\ 
 139&  17 19 55.06&  -19 41 45.15&   9.46$\pm$ 0.19&  4089\\ 
 140&  17 13 41.60&  -18 58 09.32&  11.01$\pm$ 0.28&  4038\\ 
 141&  17 25 00.72&  -20 21 26.23&  11.25$\pm$ 0.27&  4135\\ 
 142&  17 22 32.10&  -20 03 47.70&  11.05$\pm$ 0.29&  4116\\ 
 143&  17 16 25.51&  -19 19 06.63&  12.10$\pm$ 0.38&  4058\\ 
 144&  17 08 21.83&  -18 24 01.73&   9.19$\pm$ 0.19&  4009\\ 
 145&  17 23 14.53&  -20 13 19.36&  10.75$\pm$ 0.23&  4122\\ 
 146&  17 36 06.38&  -21 42 43.63&  11.29$\pm$ 0.36&  4230\\ 
 147&  17 10 42.29&  -18 45 34.57&   9.93$\pm$ 0.21&  4023\\ 
 148&  17 20 49.18&  -20 01 34.49&  11.36$\pm$ 0.30&  4098\\ 
 149&  17 14 43.31&  -19 24 38.14&  10.24$\pm$ 0.21&  4043\\ 
 150&  17 32 51.56&  -21 33 35.12&  11.49$\pm$ 0.35&  4197\\ 
 151&  17 25 49.21&  -20 46 24.36&   9.15$\pm$ 0.18&  4145\\ 
 152&  17 15 35.36&  -19 32 12.73&  10.64$\pm$ 0.25&  4050\\ 
 153&  17 37 35.13&  -22 05 34.82&  11.15$\pm$ 0.38&  4239\\ 
 154&  17 35 24.89&  -22 03 20.65&   6.09$\pm$ 0.16&  4222\\ 
 155&  17 18 28.74&  -19 57 33.35&  10.18$\pm$ 0.20&  4073\\ 
 156&  17 14 59.37&  -19 31 47.80&  10.30$\pm$ 0.22&  4046\\ 
 157&  17 25 23.64&  -20 50 27.30&   9.06$\pm$ 0.18&  4137\\ 
 158&  17 31 03.17&  -21 28 15.35&   9.75$\pm$ 0.19&  4181\\ 
 159&  17 21 48.13&  -20 24 22.27&  12.19$\pm$ 0.44&  4108\\ 
 160&  17 33 10.00&  -21 44 42.80&  11.19$\pm$ 0.29&  4198\\ 
 161&  17 20 51.98&  -20 20 29.30&  10.24$\pm$ 0.21&  4099\\ 
 162&  17 21 01.55&  -20 26 46.62&  10.58$\pm$ 0.22&  4102\\ 
 163&  17 17 53.39&  -20 09 38.39&  11.77$\pm$ 0.37&  4068\\ 
 164&  17 22 00.19&  -20 43 14.02&   9.51$\pm$ 0.18&  4109\\ 
 165&  17 36 34.05&  -22 26 57.86&   7.32$\pm$ 0.18&  4234\\ 
 166&  17 10 16.94&  -19 26 45.91&   5.65$\pm$ 0.16&  4020\\ 
 167&  17 38 11.86&  -22 35 49.08&  11.06$\pm$ 0.40&  4224\\ 
 168&  17 34 15.16&  -22 12 58.31&   8.80$\pm$ 0.18&  4210\\ 
 169&  17 21 26.92&  -20 43 49.41&  10.33$\pm$ 0.22&  4105\\ 
 170&  17 32 31.95&  -22 01 55.86&  11.54$\pm$ 0.34&  4194\\ 
 171&  17 24 06.80&  -21 05 01.94&  11.00$\pm$ 0.29&  4128\\ 
 172&  17 18 49.88&  -20 29 59.76&  11.66$\pm$ 0.34&  4079\\ 
 173&  17 36 05.96&  -22 32 38.50&   7.30$\pm$ 0.18&  4231\\ 
 174&  17 31 37.40&  -22 02 11.91&  10.63$\pm$ 0.25&  4187\\ 
 175&  17 17 51.09&  -20 26 54.97&   9.16$\pm$ 0.18&  4067\\ 
 176&  17 25 55.64&  -21 24 19.83&  10.75$\pm$ 0.25&  4146\\ 
 177&  17 15 18.48&  -20 17 00.77&  10.31$\pm$ 0.22&  4047\\ 
 178&  17 22 21.23&  -21 06 47.73&  11.25$\pm$ 0.30&  4112\\ 
 179&  17 19 11.62&  -20 46 00.91&  10.63$\pm$ 0.22&  4081\\ 
 180&  17 25 40.30&  -21 36 33.04&  10.48$\pm$ 0.24&  4141\\ 
 181&  17 20 56.90&  -21 05 24.68&   7.83$\pm$ 0.18&  4101\\ 
  \end{tabular}
\end {table*}
%&
\begin{table*}
 \begin{tabular}{@{}rrrrc@{}}
 Id. no. & $\alpha$$_{2000}$ &$\delta$$_{2000}$ & m$_{UV}$ & FAUST no. \\
\\
 182&  17 18 32.02&  -20 47 23.15&  10.49$\pm$ 0.24&  4074\\ 
 183&  17 08 54.26&  -19 40 14.29&  10.01$\pm$ 0.21&  4013\\ 
 184&  17 33 37.03&  -22 37 43.09&  11.33$\pm$ 0.34&  4203\\ 
 185&  17 19 17.46&  -21 03 02.31&  11.08$\pm$ 0.34&  4083\\ 
 186&  17 17 23.82&  -20 54 48.14&  10.01$\pm$ 0.21&  4064\\ 
 187&  17 21 59.79&  -21 24 45.48&  12.35$\pm$ 0.47&  4110\\ 
 188&  17 18 12.70&  -21 00 14.94&  10.68$\pm$ 0.27&  4069\\ 
 189&  17 20 42.84&  -21 17 32.00&  12.96$\pm$ 1.15&  4097\\ 
 190&  17 11 21.78&  -20 24 27.24&   6.92$\pm$ 0.17&  4025\\ 
 191&  17 33 53.67&  -22 58 49.90&  10.00$\pm$ 0.22&  4205\\ 
 192&  17 17 00.99&  -21 08 15.18&  10.44$\pm$ 0.26&  4062\\ 
 193&  17 06 56.00&  -19 56 57.68&  10.08$\pm$ 0.25&  4007\\ 
 194&  17 12 11.13&  -20 36 02.94&  11.16$\pm$ 0.29&  4029\\ 
 195&  17 08 32.94&  -20 12 54.43&   8.01$\pm$ 0.18&  4012\\ 
 196&  17 34 13.41&  -23 13 25.58&  11.21$\pm$ 0.36&  4208\\ 
 197&  17 14 53.61&  -21 04 56.38&   9.33$\pm$ 0.19&  4044\\ 
 198&  17 09 45.60&  -20 23 09.08&  12.08$\pm$ 0.55&  4017\\ 
 199&  17 18 36.17&  -21 34 28.09&  10.56$\pm$ 0.23&  4075\\ 
 200&  17 30 37.20&  -23 02 51.40&  10.95$\pm$ 0.29&  4177\\ 
 201&  17 31 06.05&  -23 10 38.63&   8.83$\pm$ 0.18&  4182\\ 
 202&  17 10 05.87&  -20 45 57.25&   9.45$\pm$ 0.19&  4019\\ 
 203&  17 22 53.11&  -22 20 13.20&  10.89$\pm$ 0.32&  4119\\ 
 204&  17 28 45.44&  -22 56 16.96&  11.64$\pm$ 0.40&  4163\\ 
 205&  17 19 48.48&  -21 56 20.03&  11.76$\pm$ 0.37&  4085\\ 
 206&  17 18 43.12&  -21 52 17.61&  10.93$\pm$ 0.25&  4076\\ 
 207&  17 23 52.66&  -22 36 25.83&  11.13$\pm$ 0.42&   *\\   
 208&  17 06 35.29&  -20 42 32.99&   9.78$\pm$ 0.21&  4006\\ 
 209&  17 22 35.07&  -22 41 22.11&  10.51$\pm$ 0.24&  4117\\ 
 210&  17 12 38.07&  -21 34 16.36&  10.13$\pm$ 0.25&  4034\\ 
 211&  17 18 16.43&  -22 13 47.84&  11.70$\pm$ 0.39&  4070\\ 
 212&  17 23 12.47&  -23 00 26.34&   6.84$\pm$ 0.18&  4121\\ 
 213&  17 31 03.06&  -23 52 35.98&   8.82$\pm$ 0.19&  4180\\ 
 214&  17 08 29.17&  -21 19 56.64&  10.98$\pm$ 0.29&  4110\\ 
 215&  17 18 46.62&  -22 41 00.90&  11.65$\pm$ 0.45&  4078\\ 
 216&  17 11 23.52&  -21 52 22.69&  11.82$\pm$ 0.50&  4026\\ 
 217&  17 10 43.65&  -21 59 00.33&  11.23$\pm$ 0.33&  4022\\ 
 218&  17 18 24.85&  -22 55 54.87&  11.81$\pm$ 0.52&  4072\\ 
 219&  17 24 26.09&  -23 41 52.78&  10.99$\pm$ 0.36&  4129\\ 
 220&  17 15 11.33&  -22 40 21.19&  12.33$\pm$ 0.58&   *\\   
 221&  17 24 52.83&  -23 56 02.37&   8.19$\pm$ 0.22&  4134\\ 
 222&  17 26 26.06&  -24 11 06.47&   6.89$\pm$ 0.19&  4148\\ 
 223&  17 09 07.14&  -22 13 19.09&  11.33$\pm$ 0.46&  *\\    
 224&  17 12 24.53&  -22 55 30.89&  10.03$\pm$ 0.27&  4031\\ 
 225&  17 25 14.24&  -24 37 08.66&   7.39$\pm$ 0.20&  4136\\ 
 226&  17 18 55.17&  -23 59 44.35&  11.16$\pm$ 0.49&  4080\\ 
 227&  17 20 40.36&  -24 16 08.22&   8.98$\pm$ 0.23&  4096\\ 
 228&  17 23 57.81&  -24 51 04.17&   6.31$\pm$ 0.18&   * \\
\\
  \end{tabular}
% \end{tabular}
\end{table*}
%\clearpage

\section{Identifications of the FAUST sources}
The first approach to the identification work has been the search
for positional coincidence with various catalogues such as the 
Smithsonian Astrophysical Observatory (SAO 1966) 
and the Hipparcos  Input Catalog (Turon \etal 1993), 
mainly through the SIMBAD data base. 
In addition to the positional coincidence (3'x3' error box), some astrophysical 
constraints have been applied, such as that when more than one 
counterpart lies in the error box of a UV source, the bluest or  brightest  or
earlier-type star was accepted. This search led to a positive identification
of 166 sources($\simeq$ 73 \%). 

For the remaining  sources no plausible counterpart could be found using 
correlations with catalogues entries.
This sky region is very crowded, owing to the low galactic latitude of the
FAUST frame, and a visual inspection of the E and O prints of the 
Palomar Sky Survey (PSS) does not allow one to select a reasonable number of
candidates among the many stars inside the error box of the UV source.

The adopted approach was to first narrow down the list of possible
objects by obtaining multicolor broad-band photometry of the 
region around each of the unidentified UV sources. The color--color
diagram for all the stars inside the error box was used to select
the bluest objects, and low-resolution spectra were obtained
for these in order to assign a most probable counterpart to the FAUST source. 

\section{Ground-based observations}

The search for an optical counterpart to the unidentified FAUST
UV sources was performed with  ground-based photometry and
spectroscopy obtained at the Wise Observatory.

\subsection {Multicolor photometry}

For each of the 62 unidentified FAUST sources, three CCD short exposures 
have been obtained in blue, visual and red (B, V, R) at the 1m telescope 
of the Wise Observatory. 
The camera used is a Tektronix thinned and back-illuminated 1024 x 1024 
pixel  CCD, which  images a field of view of 12'.
% (Kaspi et al. 1995).
 
Typical exposure times were 180 sec for B, 90 sec for V,
and 80 sec for R. Photometric reduction was performed using the DAOPHOT
package, and the instrumental magnitudes and color indices 
(b--v) and (v--r) were obtained for all the stars inside a  3'x3'  
error box around each UV source. 

 The mean intrinsic colors of a star of a given spectral type and 
luminosity class are well-defined (Johnson 1966). This implies that a 
color-color diagram can be used to select stars of different 
spectral types. By plotting the instrumental color indices for all the
objects within the error box, the bluest (or earliest-type) stars should
reside in a well-defined part of the diagram.
    
However, the mean colors are well-determined only for unreddened stars.
% i.e.,after removing the effects of interstellar extinction. The question is 
Hence we must address the question on how reddening may affect the 
measured colors  and if the use of 
reddened colors could bias our selection procedure, which is based on 
the (b--v) and (v--r) colors. In the following, we assume that there are 
no significant color terms in the transformation from instrumental to 
standard colors. This is justified by numerous absolute-photometry 
studies performed at the Wise Observatory with CCD cameras.

We  plot  in Fig 3 the mean colors (B--V)$_{0}$ versus
(V--R)$_{0}$ for main-sequence, giant and supergiant stars 
(Johnson 1966). Note that the intrinsic (B--V)$_{0}$ of a B0 type main 
sequence star is  -0.30  while it is (B--V)$_{0}$= +0.82 for a K0 star.
It can be seen that, up to (B--V)$_{0}$ $\simeq $ 1.0, the locus of the
 intrinsic colors of the stars is well-defined as  a linear relation with 
an approximate
slope of 0.80 for all the luminosity classes. The lines representing the 
supergiant and bright giant stars diverge from the main sequence line
only for the very red stars with  (B-V)$>$1.0.
From the average extinction curve (Savage \& Mathis 1979),
the slope of the reddening line   E(V--R)/E(B--V) is 0.78, very similar to 
the slope of the intrinsic color-color relation. Hence, the use of 
instrumental colors instead of intrinsic ones will only  
underestimate the spectral type along the reddening line. This
increases our confidence that the use of the two-color diagram (v--r) 
versus (b--v) is a suitable tool for selecting the blue color-excess
objects inside the error box of the FAUST source.
\begin{figure}
\hspace{10cm}
\centerline{\epsfxsize=3.5in\epsfbox{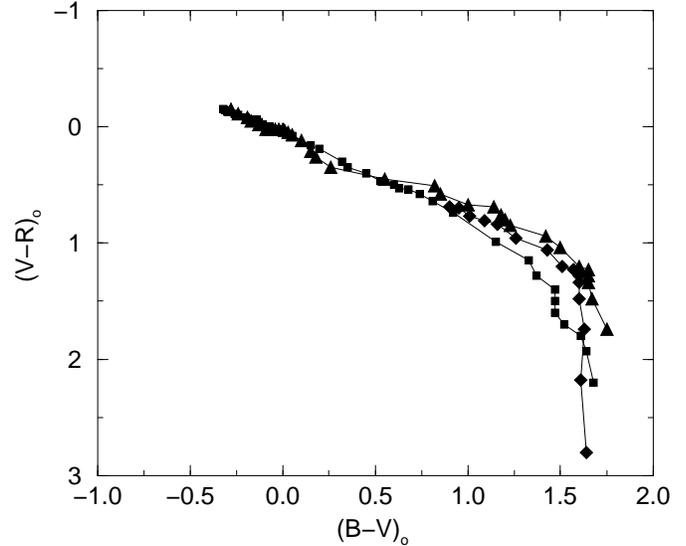}}
\caption{(V-R)$_{0}$ versus (B-V)$_{0}$ colors diagram for stars. Main sequence
 stars are plotted as squares, giant stars  as diamonds and 
supergiants are plotted as triangles.}
\end{figure}
 
Fig 4 shows the PSS image of the field around the UV source N.92, 
whose location is marked by a cross.
The brightest star in the field is HD 156169, classified as K0 III. 
\begin{figure}
%\hspace{14cm}
\centerline{\epsfxsize=3.0in\epsfbox{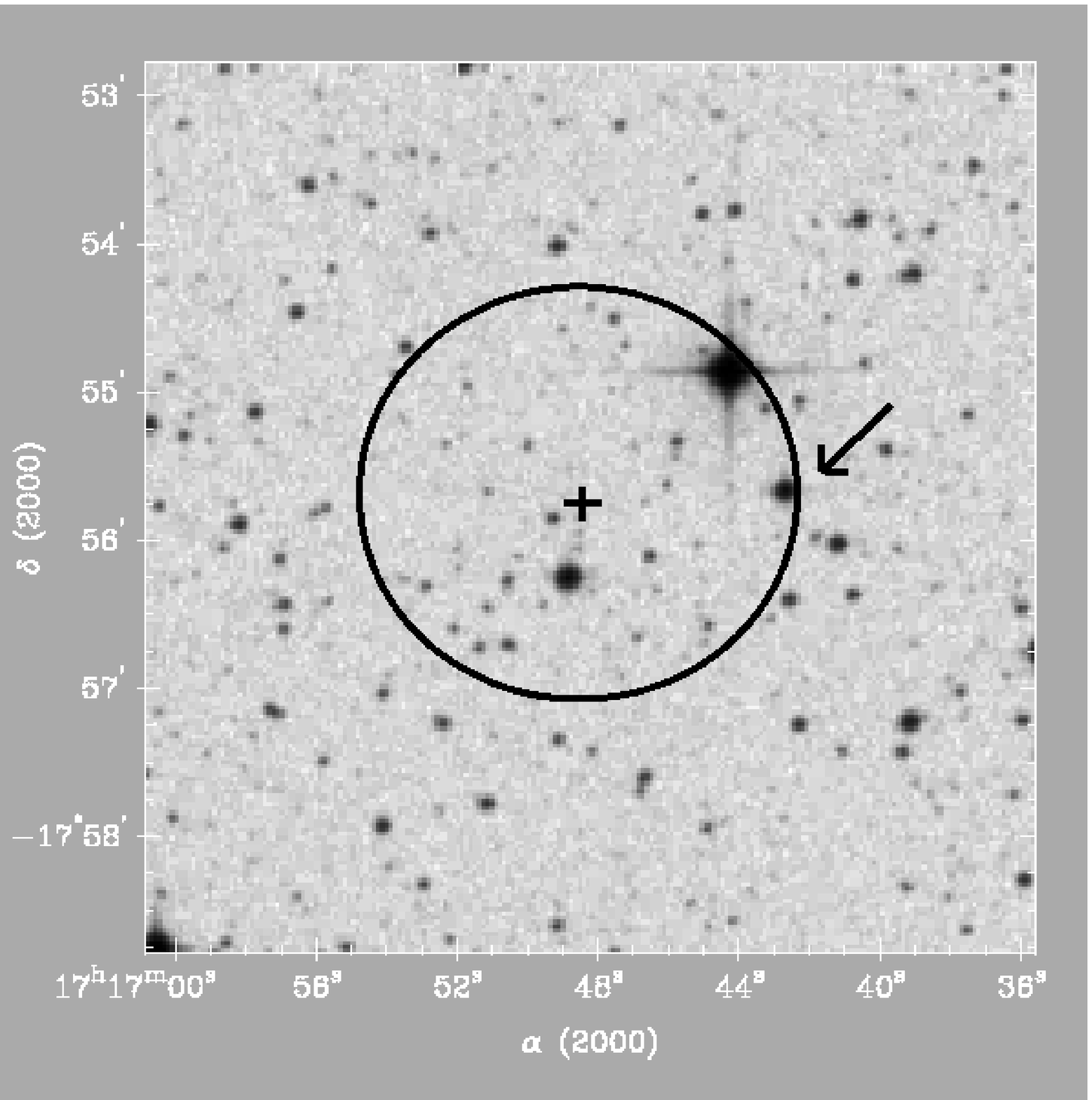}}
%\centerline{\epsfxsize=2.5in\epsfbox{fig4.eps}}
\caption{PSS  segment of the field around UV source N. 92.
The position of the FAUST source is marked by a cross, and the optical
counterpart by an arrow. The circle indicates the site of the positional 
error expected for this Faust source.}
\end{figure}

Three short CCD exposures in B, V, and R 
were obtained for this field and blue, visual and red instrumental magnitudes 
were measured for all the objects in a 3'x3' box centered on the cross.
The  (b--v) versus (v--r) diagram is
shown in Fig 5. It can be seen that all  the measured objects, 
including the one marked by an asterisk, lie on a well defined range of 
color indices.
This object, characterized by the bluest instrumental (b--v) color,
was picked up as the possible counterpart to the UV source to be
observed spectroscopically. This star, marked by an arrow in Fig 4, 
was observed later at the Wise Observatory and classified as an 
early-A type of the proper apparent V magnitude to have the measured
UV flux. This test demonstrates the validity of selecting counterparts
of UV sources based on color-color diagrams.

\begin{figure}
\hspace{10cm}
\centerline{\epsfxsize=3.0in\epsfbox{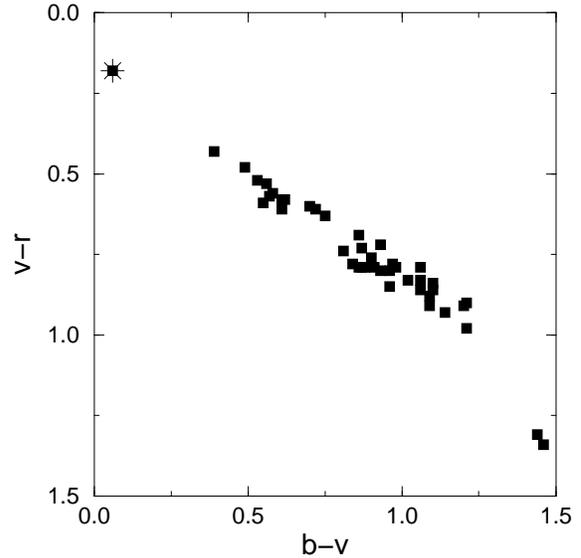}}
\caption{The two color diagram (v-r) versus (b-v) for the stars
measured in a three arcmin box around the N. 92 source. The asterisk marks
the star chosen as candidate, further classified as an early-A star.}
\end{figure}

For all the UV sources (but one) for which we applied the photometric
method, the two-color diagram allowed the selection  of one or a few
candidates with  a blue color excess.

The magnitudes and colors used for the selection are instrumental. Our color
calibration is poor, since  photometric standard stars were obtained
only on a few of the observation nights. We compared the Wise calibrated
magnitudes with the $m_{v} $ magnitude from the Guide Star Catalog (Lasker et al. 1987) 
and found a systematic difference of 0.1 mag for our  visual magnitude; this 
is quite acceptable and does not affect the selection criteria.

Fig. 6 shows the calibrated V--R versus  B--V for the stars measured at Wise.
The colors for main sequence  stars of known spectral type (Johnson 1966)
are also marked.
It is evident that if we correct for the systematic shift in our magnitudes,
and apply the 0.1 mag shift to the visual magnitude, the measured points will lie on the 
the photoelectric relation. 
\begin{figure}
%\hspace{10cm}
%\centerline{\epsfxsize=3.5in\epsfbox{fig6.eps}}
\centerline{\epsfxsize=4.0in\epsfbox{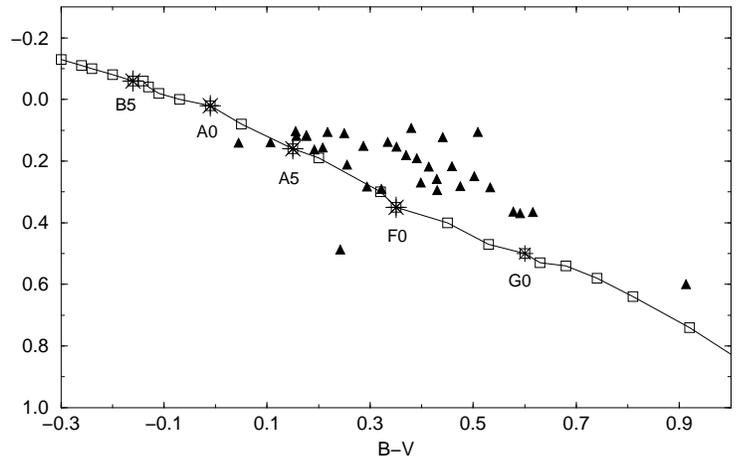}}
\caption{The (V-R) versus (B-V) diagram for the photometrically
 observed stars. Triangles represent the stars  for which photometric
 calibration was available. Squares indicates the main sequence stars
(Johnson et al. 1966).}
\end{figure}

\subsection {Spectroscopy}

The next step was to obtain spectra of the color-excess candidates, selected
by the multicolor photometry.
We used the FOSC (Faint Object Spectrometer Camera) at the f/7 focal 
position of the 1.0-meter telescope at the Wise Observatory and the Tektronix
1024x1024 pixel CCD.
The wavelength coverage used was  from $\sim 4000 $\AA\ to $\sim 8000 $\AA\,
with a resolution of $\sim 4$\AA\  per pixel.  
All images were 
bias-subtracted and flat-fielded, and all spectra were extracted and
wavelength and flux-calibrated using  IRAF.

Spectral types were derived by comparing the observed spectra with those in
the spectral atlases of Jacoby et al. (1984) and of Silva \& Cornell (1992).
Spectral subclass types could not be assigned, since the spectral coverage
did not include the blue region from $\sim 3800 $\AA\  to $\sim 4000 $\AA\
where the relevant lines are located. Only a broad subclass is defined here,
such as early or late, and is marked in Table 2 with the symbols e or l.
Most spectra show only prominent Balmer absorption lines and are classified 
as  early--type stars, from mid--B to late--A or F.

The low resolution of the spectra prevents a luminosity classification,
but supergiants  appear in most cases to be ruled out, because supergiant  
B stars have almost no H$\alpha$ absorption while late-A supergiants
show reasonably deep He I  $\lambda  5876 $\AA\, absorption.

Table 2 lists the optical counterparts to the FAUST sources and their 
spectral types obtained from  SIMBAD or derived from our observations. 
The spectroscopic observations obtained at Wise allowed us to assign a 
spectral type to 60  stars among the 62 unidentified sources.

\begin{table*}
\caption{Identified FAUST sources}
\begin{tabular}{@{}rccccccrc@{}}
Id no.& $\alpha$$_{2000}$ & $\delta$$_{2000}$& Proposed id. & Sp. Type & m$_{b}$ & m$_{v}$\\
\\
   1&17 34 46.98& -16 37 25.9&  SAO 160650         & B9II/III        &   8.76  &   8.74  \\
   2&17 29 53.77& -16 00 57.5&  SAO 160582         & A0IV            &   8.62  &   8.48  \\
   3&17 30 57.66& -16 09 53.6&  SAO 160605         & A2/A3IV         &  9.77  &  9.42    \\
   4&17 28 55.86& -16 01 10.1&  SAO 160569         & A0V             &  10.13 &   9.95   \\
   5&17 32 32.40& -16 38 34.4&   HD 158924         & B9II/III        &   9.54 &   9.40   \\
   6&17 29 18.35& -16 35 26.7&   SAO 160574        & B8Vne           & 8.92  &   8.894   \\
   7&17 34 05.76& -17 08 18.2&   SAO 160644         & B9II            &   9.96 &   9.83  \\
   8&17 33 37.64& -17 06 34.3&   HD 159123         & B9IV            &  10.43 & 10.13    \\
   9&17 23 53.00& -16 02 11.0& wise(172353-160211) & A(e)            &  11.2  &          \\
  10&17 31 12.83& -17 08 31.7&   SAO 160608         & B0II            & 8.32 & 8.20      \\
  11&17 37 39.10& -17 49 76.1&   SAO 160702         & B1Ib            & 8.58 & 8.56      \\
  12&17 25 35.48& -16 26 48.1&   SAO 160536         & A3III           &   9.9  & 9.57    \\
  13&17 25 33.00& -16 35 40.0& wise(172533-163540) & B(l)            &  11.4  &          \\
  14&17 35 33.00& -17 49 33.0& wise(173533-174933) & B(e)            &  11.2  &          \\
  15&17 23 21.00& -16 28 27.0& wise(172321-162827) & A(e)            &  10.8  &          \\
  16&17 18 39.22& -16 01 47.1&   SAO 160445          &A3III          &  7.78  & 7.61     \\
  17&17 39 18.01& -18 24 41.4&   SAO 160720         &B1/B2Ib/II      &9.17    &9.07      \\
  18&17 33 58.99& -17 50 10.6&   SAO 160642         & F2V            &   8.27 & 7.93     \\
  19&17 28 39.29& -17 14 58.2&   HD  158195        &B2III            &  10.18 &10.11     \\
  20&17 30 30.00& -17 29 09.0& wise(173030-172909) &A(e)             &  11.2  &          \\
  21&17 35 11.00& -18 02 12.0& wise(173511-180212) & B(e)            &  10.6  &          \\
  22&17 26 49.00& -17 06 10.0& wise(172649-170610) & B(e)            &  11.2  & 11.1     \\
  23&17 34 41.00& -18 04 53.0& wise(173441-180453) & A(e)            &  10.5  &          \\
  24&17 32 14.85& -17 46 54.8& HD  158841          & A0III           &  11.4  &  11.3    \\
  25&17 28 51.28& -17 23 21.9& HD  158231          & B9III           &  10.64 &  10.5    \\
  26&17 29 44.79& -17 35 34.1& SAO 160581          &  A1V            &   9.14 &  8.97    \\
  27&17 35 20.04& -18 14 42.9& SAO 160661          &B9.5/A0V         &   9.33 &  9.17    \\
  28&17 25 26.00& -17 05 25.0& wise(172526-170525) &A(e)             &        &  11.9    \\
  29&17 36 18.40& -18 24 23.8& HD 159590           &B9.5V            &  10.23 &  10.1    \\
  30&17 39 08.00& -18 43 13.8& HD 160144           & B9III/IV        &  10.21 &  10.2    \\
  31&17 33 19.43& -18 02 33.5& SAO 160632          &A1III/IV         &   9.69 &   9.32   \\
  32&17 34 33.39& -18 13 15.0&   SAO 160649        &F0IV/V           &   7.49 &  7.11    \\
  33&17 21 08.00& -16 38 38.0& wise(172108-163838)  & A(e)           &  12.4  &          \\
  34&17 16 18.42& -16 05 48.0&  SAO 160414         & A0V             &   9.32 & 9.06     \\
  35&17 35 01.97& -18 19 39.1&      SAO 160658     & A0IV            &  9.97  & 9.84     \\
  36&17 32 19.00& -18 02 39.0&wise(173210-180239)  &  B             &  10.7   &          \\
  37&17 20 02.55& -16 40 55.7&   SAO 160467        &B9IV            &   8.76  & 8.62     \\
  38&17 19 33.00& -16 32 42.0& wise(171933-163242) & A(e)           &  11.2   &          \\
  39&17 35 55.22& -18 34 00.4&      HD 159506      &A0V             &  9.79   &  9.64    \\
  40&17 14 32.89& -15 57 22.9&     SAO 160395      & A0             &   8.81  & 8.66     \\ 
  41&17 16 11.98& -16 15 25.1&      SAO 160412     & B9V            &   9.93  & 9.83     \\
  42&17 30 18.58& -17 56 26.7&  wise(173018-175627) & no sp         & 10.81   &10.4      \\
    &17 30 00.58& -17 55 05.7&     HD 158442       & A2             &  10.35& 10.11      \\
  43&17 39 28.19& -19 01 07.3&    HD  160219       & B9IV           &   10.43   &10.2    \\
  44&17 33 27.00& -18 20 23.2&   SAO 160635        & F3V            &  8.20    & 7.76    \\
  45&17 26 31.65& -17 35 13.5&   HD 157859         & A2V            &  10.26  & 9.92     \\
  46&17 32 22.00& -18 15 15.0& wise(173222-181515) & A(l)           &  11.1 &            \\
  47&17 34 21.10& -18 36 21.7&   HD 159246         & B9III          &  9.96 & 9.81       \\
  48&17 18 14.25& -16 40 48.3&  SAO 160436         &A0IV            &   9.52 &  9.35     \\
  49&17 33 25.00& -18 31 16.0& wise(173325-183116) & A(e)           &  11.3  & 10.8      \\
  50&17 38 21.82& -19 04 26.8&   HD 159976         & A0IV           &  10.74  & 10.5     \\
  51&17 22 01.75& -17 13 11.1&   SAO 160496        & A5III          &   8.68& 8.42       \\
  52&17 15 39.00& -16 27 54.0& wise(171539-162754) & B              &  12.4 &            \\
  53&17 22 22.56& -17 20 25.6&   SAO 160500        & A0V            &  8.10  & 7.97      \\
  54&17 29 44.42& -18 07 47.1&   HD  158391        & A8V            &  10.66 &10.4       \\
  55&17 34 17.64& -18 46 36.2&   SAO 160647        & B8II/III       & 9.79   &9.70       \\
  56&17 28 13.00& -18 00 24.0& wise(172813-180024) & A(l)           &  11.2  &           \\
  57&17 26 44.15& -17 52 34.6&  HD 157897          & A0IV/V         & 9.94   &  9.84     \\
  58&17 20 25.63& -17 13 15.0&   HD  156830        & B9III          &10.72   & 10.6      \\
\end{tabular}
\end{table*}
\begin{table*}
\begin{tabular}{@{}rccccccrc@{}}
Id no.& $\alpha$$_{2000}$ & $\delta$$_{2000}$& Proposed id. & Sp. Type & m$_{b}$ & m$_{v}$\\
 \\
  59&17 20 52.71& -17 20 05.2&  V441 OPH           & A0             &  11.5  &  11.8     \\
  60&17 26 31.02& -18 05 07.4&  HD  157860         & B9III          & 10.46  & 10.27     \\
  61&17 37 00.00& -19 20 22.0& wise(173700-192022) &A(e)            &  10.6  &           \\
  62&17 32 32.00& -18 50 03.0& wise(173232-185003) &A(e)            &  11.4  &10.4       \\
  63&17 19 40.66& -17 39 23.4&  SAO 160505         & A1IV/V         &  8.35  & 8.22      \\
  64&17 12 58.58& -16 30 52.6&  SAO 160368         &  A1m           &  9.06  &  8.77     \\
  65&17 29 44.00& -18 32 58.0&  wise(172944-183258)& A(l)           &  10.7  &           \\
  66&17 26 41.91& -18 15 26.5&  wise(172642-181526)&A               &  10.8  &           \\
  67&17 31 35.00& -18 50 11.0&  wise(173135-185011)& A(l)           &  11.2  &           \\
  68&17 26 30.67& -18 37 31.5&  SAO 160576         & Ap             &  9.46  & 9.35      \\
  69&17 23 25.40& -17 58 15.7&  SAO 160513         &A2III/IV        & 8.81   & 8.59      \\
  70&17 11 29.57& -16 29 30.8&  SAO 160347         & A7V            & 7.78   &  7.52     \\
  71&17 25 20.62& -18 16 53.4&   SAO 160533              &A0V       & 8.73   &  8.58     \\
  72&17 32 21.00& -19 03 28.0&   wise(173221-190328)   & A(l)       &  10.7 &            \\
  73&17 22 18.00& -17 53 15.0&   wise(172218-175315)    &  A(l)     &  11.3 & 11.3       \\
  74&17 28 08.00& -18 36 23.0&   wise(172808-183623)     & A(l)     &  10.4  &           \\
  75&17 19 53.35& -17 45 23.6&   SAO 160462            &A0V         &  6.06   & 6.04     \\
  76&17 33 15.10& -19 22 37.9&   SAO 160630            & B9IV       & 9.74    &   9.52   \\
  77&17 16 11.00& -17 19 47.0&     wise(171611-171947)  &A(e)       & 11.4    &          \\
  78&17 21 15.86& -18 02 54.4&     SAO 160487           & B9.5V     & 9.45    & 9.40     \\
  79&17 24 37.04& -18 26 44.7&     SAO 160523           & B8V       &  6.38  &  6.21     \\
  80&17 22 50.51& -18 11 19.9&     SAO 160506            &A7III/IV   & 9.98   &    9.66  \\
  81&17 37 32.10& -20 03 15.4&     SAO 185560            & A0V  & 10.2&  9.1             \\
  82&17 17 01.95& -17 39 25.1&     SAO 160422            &  A0V &  9.03  &  8.86         \\
  83&17 35 33.30& -19 53 48.2&     HD  159432            &  A0V & 9.1 & 8.1              \\
  84&17 23 24.00& -18 26 06.0&    wise(172324-182606)    & A(e) & 11.3 &                 \\
  85&17 17 33.10& -17 52 13.7&     SAO 160429            &  B9V&  9.46 &9.41             \\
  86&17 14 21.67& -17 27 57.9&     SAO 160393            & B9III/IV& 8.20   &  8.15      \\
  87&17 12 15.33& -17 14 47.5&     SAO 160354            &AIV  &  7.62  & 7.57           \\
  88&17 27 02.69& -18 59 31.1&     SAO 160551            & B9Ib/II& 9.84 & 9.48   \\
  89&17 38 33.00& -20 17 28.0&     wise(173833-201728)   & A(l) & 10.4  &     \\
  90&17 24 28.57& -18 50 49.8&     SAO 160522            &    A0V & 9.20 &9.04    \\
  91&17 30 38.78& -19 33 27.0&     SAO 160595            & A2III   & 9.3& 8.1    \\
  92&17 16 44.13& -17 55 41.6&     wise(171644-175542)   & A(e)   &  11.4 &      \\
  93&17 27 51.28& -19 17 07.3&     HD  158052            & B9V &  10.56& 10.16      \\
  94&17 39 17.88& -20 42 32.4&     SAO 185596                  &  B9V &9.41 &  9.3    \\
  95&17 34 23.61& -20 06 11.3&     SAO 185516            & A0III/IV& 10.24    &10.04     \\
  96&17 15 31.00& -17 50 54.0&    wise(171531-175054)    &  G   & 12.7  &     \\
  97&17 19 43.00& -18 21 54.0&    wise(171943-182154)    &  no sp  &$>$15 &      \\
  98&17 10 20.17& -17 19 44.6&     HD  155108               &  A0III&9.96   &9.77     \\
  99&17 12 28.74& -17 36 13.0&     SAO 160359             & A9V  & 8.69  &     8.35      \\
 100&17 25 47.96& -19 15 05.9&     HD  157732              & A0/A1V&10.16  &9.83      \\
 101&17 37 50.53& -20 40 03.5&     HD  159880                 & B8/B9V&   9.60 & 9.48     \\
 102&17 16 06.00& -18 31 00.0&     HD  156587            & nebula     &  &            \\
 103&17 35 46.09& -20 27 59.1& wise(173546-202759)              & A(e)& 10.8  &      \\
 104&17 15 44.20& -18 10 54.0&     HD  155992               &A3II/III& 9.4 &   9.3       \\
 105&17 20 15.74& -18 49 24.9&     SAO 160469              &   B2II&  9.40&   9.29       \\
 106&17 26 23.96& -19 31 04.3&    SAO 160542                & A0 V       &  9.63 & 9.32   \\
 107&17 38 45.61& -20 57 37.6&     SAO 185585               &  B8Ib/II   &  9.64 & 9.49    \\
 108&17 20 53.57& -18 54 12.1&     HD  156895               &  B8III     & 10.45 & 10.37   \\
 109&17 22 30.14& -19 06 09.9&     SAO 160501                &  B9.5V    &  9.76 &  9.54  \\
 110&17 35 21.46& -20 39 28.7&     HD  159394                &  A0IV     & 10.42 & 10.3   \\
 111&17 14 27.65& -18 12 24.6&     HD  155791                &  A0IV     & 10.8  & 10.6      \\
 112&17 09 04.51& -17 34 09.7&     SAO 160315                &  A0V      &  8.43 &  8.33     \\
 113&17 34 13.00& -20 33 44.0&    wise(173413-203344)        &  A     & 11.0  &             \\
 114&17 13 54.58& -18 13 07.6&      SAO 160385               &  A9V      &  8.56 & 8.3       \\
 115&17 36 46.25& -20 57 22.5&     SAO 185548                &  A2V     &  8.46  & 8.06      \\
 116&17 35 39.35& -20 48 35.2&     SAO 185535                &  A1III   &  9.14  &  8.88     \\
 117&17 15 29.18& -18 25 37.6&     HD  155957                &  A2/3II  & 10.57  &   10.27   \\
 118&17 12 29.39& -18 08 12.8&     SAO 160358                &  B8II    &  9.12  &    9.12   \\
 119&17 15 04.00& -18 26 00.0&     wise(171504-182600)       &  A(e)    & 11.7   &           \\
 120&17 23 16.16& -19 26 55.4&     SAO 160511                &  A2III   &  8.43  & 8.04      \\
 121&17 40 14.70& -21 21 29.7&     wise(174015-212130)       &  B       & 11.4   &11.3       \\
 122&17 25 29.26& -19 44 27.1&     SAO 160534                &  F0V     & 8.54   & 8.26      \\
 123&17 37 33.64& -21 10 22.3&     wise(173734-211022)       &  A(e)    & 11.2   & 10.8      \\
\end{tabular}
\end{table*}
\begin{table*}
\begin{tabular}{@{}rccccccrc@{}}
Id no.& $\alpha$$_{2000}$ & $\delta$$_{2000}$& Proposed id. & Sp. Type & m$_{b}$ & m$_{v}$\\
\\
 124&17 26 48.12& -19 59 15.8&     HD  157898                & A3II/III & 10.2   & 9.1       \\
 125&17 11 06.26& -18 09 30.0&     SAO 160341                & A0IV/V  &  10.28  & 10.17     \\
 126&17 34 48.00& -21 03 00.0&     HD  159290                   & B9V  & 11.8    & 10.8      \\
 127&17 27 13.90& -20 18 52.6&     SAO 185414                   & B9III& 9.34    &  9.19     \\
 128&17 24 39.00& -19 57 08.0&     wise(172439-195708)          &  A(e) & 11.3    &          \\
 129&17 17 29.82& -19 09 10.7&     HD 156289                   &B9.5II &11.1     &  10.9     \\
 130&17 12 40.45& -18 33 19.9&     SAO 160363                   & A0II  & 9.46    &  9.17    \\
 131&17 11 57.65& -18 28 00.4&     HD 155377                    & B9IV/V& 11.0    &  10.7     \\
 132&17 37 19.99& -21 32 20.8&     SAO 185555                   & B9IV  &  9.84   &   9.64    \\
 133&17 08 27.05& -18 10 12.3&     SAO 160309                   & B9V  &   8.38    &   8.33  \\
 134&17 36 46.00& -21 32 35.0&     wise(173646-213235)          &  A(e)&  10.8    &          \\
 135&17 36 16.78& -21 28 05.5&     wise(173617-212806)          &  A(l) & 11.4   & 11.2     \\
 136&17 27 14.70& -20 27 51.4&     HD  157954                   &  B9  & 10.7    &  10.7     \\
 137&17 09 07.52& -18 15 22.8&     SAO 160317                   &  A7III &  8.79 &  8.49     \\
 138&17 29 02.74& -20 45 21.6&     wise(172903-204522)          &  B(e)  & 11.4 &  10.9       \\
 139&17 19 56.31& -19 42 15.6&     HD  156719                   &A3III &  8.24 & 8.00        \\
 140&17 13 41.15& -18 58 39.0&     SAO 160381             & A9V   & 8.85 & 8.57              \\
 141&17 25 31.77& -20 19 10.2&     SAO185386              &A4II   & 9.91 & 9.56               \\
 142&17 22 34.20& -20 04 09.9&     HD  157185             &   B9V       & 10.52 & 10.27         \\
 143&17 16 28.64& -19 20 14.3&     HD  156095             & A9V         & 8.97  & 8.53       \\
 144&17 08 19.08& -18 22 41.5&     SAO 160307             &    A1IV     & 8.96  & 8.79       \\
 145&17 23 17.16& -20 13 49.4&     HD  157312             &  B9III      & 10.34 & 10.16       \\
 146&17 36 00.00& -21 42 25.0&   wise(173600-214225)      &    A(l)     & 10.6  &      \\
 147&17 10 41.36& -18 45 17.7&     SAO 160334             & B9II/III    &  10.0 &      \\
 148&17 20 51.00& -20 02 29.0&     wise(172051-200229)    &  A(l),F     & 11.4  &            \\
 149&17 14 44.00& -19 24 57.0&     wise(171444-192457)    & A        & 11.4  &      \\
 150&17 32 50.64& -21 34 18.2&     HD  158939             & A0V         & 10.76 & 10.5      \\
 151&17 25 51.45& -20 46 48.3&     SAO 185393             & A0V         & 9.17  &  8.96     \\
 152&17 15 37.00& -19 33 00.0&     wise(171537-193300)    & G(e)        & 13.1  &      \\
 153&17 37 26.91& -22 04 41.6&      HD 159781             & A0IV        & 10.36 & 10.1      \\
 154&17 35 18.50& -22 02 37.8&     SAO 185526             &   Ap        &  6.49 &6.57   \\
 155&17 18 31.60& -19 58 30.7&     HD  156490             & B9III       & 10.19 & 10.01    \\
 156&17 15 00.72& -19 32 31.8&     SAO 160399             & B9IV        & 9.70  & 9.51  \\
 157&17 25 25.89& -20 50 58.3&     SAO 185384             & B9III       & 9.70  &9.60     \\
 158&17 31 03.46& -21 29 07.0&     HD  158596             & Ap          & 9.17  &8.94   \\
 159&17 21 51.28& -20 25 03.4&     HD  157055             & A3II/III    &10.45  &10.29        \\
 160&17 33 09.82& -21 45 40.7&     SAO 185498             &  A1/A2IV    & 9.88  &    9.62       \\
 161&17 20 55.47& -20 21 35.1&     SAO 185294             &  A0V        & 10.02 & 9.89       \\
 162&17 21 06.87& -20 29 08.1&    SAO 185302              & A8/A9V      & 9.1   &   8.83     \\
 163&17 17 55.84& -20 10 28.1&    wise(171756-201028)     & A(e)        &10.89  & 10.6   \\
 164&17 22 03.96& -20 44 29.0&     SAO 185323             & B9.5IV/V    & 9.36  & 9.04    \\
 165&17 36 29.03& -22 26 27.9&     SAO 185545             & B8IV        &8.95    &8.85    \\
 166&17 10 14.94& -19 26 10.6&     SAO 160326             & B8Ib/II     &  6.945 &7.03   \\
 167&17 38 03.82& -22 34 54.9&     SAO 185572             & F5V         &  8.47  & 8.13    \\
 168&17 34 12.38& -22 13 20.5&     SAO 185515             & B8IV        & 9.53   &9.43       \\
 169&17 21 33.00& -20 44 40.0&     wise(172133-204440)    &  A(l)       & 11.5 &       \\
 170&17 32 31.24& -22 02 32.4&     HD  158858             &A2III        & 10.58   &10.25       \\
 171&17 24 09.00& -21 05 49.0&     wise(172409-210549)    & A(l)    & 10.9    &              \\
 172&17 18 52.38& -20 29 57.4&     SAO 185258             & F0V     &  8.97   & 8.62        \\
 173&17 36 02.25& -22 32 33.6&     SAO 185540             &  A0IV   & 8.69    & 8.51    \\
 174&17 31 35.30& -22 02 57.1&     SAO 185473             &  B6II   &10.56    & 10.14       \\
 175&17 17 53.78& -20 27 49.7&     SAO 185236              &B9III/IV & 9.66  & 9.60   \\
 176&17 25 57.42& -21 24 58.7&     SAO 185396              &   Ap    & 9.65  &  9.29  \\
 177&17 15 18.73& -20 17 41.2&     SAO 185201+             &   A3V   & 9.44  & 9.14 \\
    &17 15 17.69& -20 16 24.0&     HD 155927               & A8V     & 9.77  & 9.36 \\
 178&17 22 25.08& -21 08 31.8&     SAO 185332              & B9IV/V  & 10.53 &  10.3  \\
 179&17 19 23.00& -20 46 172.0&      wise(171923-204617)    & F(e)    & $>$13 &         \\
 180&17 25 43.19& -21 37 40.1&     HD 157706               & B9III   & 10.87 & 10.8       \\
 181&17 21 00.37& -21 06 46.6&     SAO 185296              &F2/F3V   & 4.78 & 4.38    \\
 182&17 18 31.41& -20 48 19.6&     SAO 185248              &   A2III &  9.66 & 9.41     \\
 183&17 08 53.41& -19 40 35.1&     SAO 160311              &  A0IV   & 9.80  & 9.69         \\
 184&17 33 33.09& -22 38 54.4&     SAO 185504              &   A6IV  & 9.49  & 9.18 \\
 185&17 19 20.65& -21 04 29.4&     HD 156601               & B9.5IV  & 10.48 & 10.3     \\
 186&17 17 25.65& -20 55 39.9&     SAO 185229              &  A3     & 10.19 &    9.92    \\
 187&17 22 04.00& -21 26 51.0& wise(172204-212651)         & B(l)    & 10.0  &     \\

\end{tabular}
\end{table*}

\begin{table*}
\begin{tabular}{@{}rccccccrc@{}}
Id no.& $\alpha$$_{2000}$ & $\delta$$_{2000}$& Proposed id. & Sp. Type & m$_{b}$ & m$_{v}$\\
\\
 188&17 18 13.75& -21 00 32.8& wise(171814-210033)         & A(e)    & 10.1  &     \\
 189&17 20 42.74& -21 18 58.7& wise(172043-211859)         & A(e)    & 10.1  &      \\
 190&17 11 21.99& -20 25 29.0&    SAO 185122               &  B9IV   & 7.50  & 7.52 \\
 191&17 33 45.31& -22 58 41.6&    SAO 185509                 & A0V  &  9.87   & 9.61   \\
 192&17 17 02.00& -21 09 08.0&  wise(171702-210908)          &  A(e)   & 10.3 &      \\
 193&17 06 56.50& -19 56 45.8&     SAO 160283                &  A0III  &  9.18  &  9.08    \\
 194&17 12 11.79& -20 37 18.7&   wise(171212-203719)         &  A(e)   & 11.1& 10.4    \\
 195&17 08 32.48& -20 13 03.1&     SAO 185067                & A0IV   &7.94     & 7.79     \\
 196&17 34 07.54& -23 13 59.1&     SAO 185513                & A1IV    & 10.27  & 9.98       \\
 197&17 14 54.89& -21 06 00.3&     HD  155847                & B5III   & 10.12  &9.94     \\
 198&17 09 44.32& -20 22 48.62&    SAO 185090                & A2II/III& 10.27  &10.05     \\
 199&17 18 38.77& -21 35 21.8&     SAO 185252                & B3III   &  9.88  &9.52   \\
 200&17 30 34.34& -23 02 14.8&   wise(173034-230213)         & A(e)    & 9.4   &   \\
 201&17 31 03.02& -23 10 30.8&     SAO 185467                & B7III   & 9.14   &    8.94    \\
 202&17 10 06.08& -20 47 01.2&     HD  155064                & B9V     & 9.57 & 9.37   \\
 203&17 22 54.62& -22 21 03.8&   wise(172255-222103)         &  B(e)   & 11.4 &11.1  \\
 204&17 28 45.00& -22 56 20.0&   wise(172845-225620)         & A(e)    & 11.0 &      \\
 205&17 19 50.42& -21 57 00.6&   wise(171950-215700)         & A(e)    & 11.3 & 11.1    \\
 206&17 18 45.81& -21 53 11.0&     HD  156503                &  A0III  & 10.83   &   10.5      \\
 207&17 23 52.66& -22 36 25.8&       wise phot  no cand      &  NO ID  & &          \\
 208&17 06 34.44& -20 42 41.3&     HD  154498                &  B9III  & 10.42 &  10.31      \\
 209&17 22 37.71& -22 42 01.8&     SAO 185333                & A1III/IV &  9.48 &  9.1    \\
 210&17 12 39.53& -21 36 21.8&     SAO 185150              &   F2V     & 7.6      & 6.8   \\
 211&17 18 21.00& -22 15 20.0&      wise(171821-221520)    &   B(e)    & 10.8     &      \\
 212&17 23 13.10& -23 00 34.4&     SAO 185349              &    B9V    &  7.46    & 7.40    \\
 213&17 30 58.02& -23 52 55.2&   wise(173058-235255)       &    A   & $>$14    &       \\
 214&17 08 28.31& -21 20 13.4&     HD 154804               &   A0IV    & 11.2     & 10.9     & \\
 215&17 18 54.79& -22 42 28.2&     SAO 185257              &   F0V     &  9.53    &9.13   \\
 216&17 11 25.52& -21 53 50.6&     SAO 185123              &   A5IV    & 9.46     &9.05    \\
 217&17 10 43.91& -22 00 16.5&     SAO 185109              &  A2/A3IV  &  9.25    & 8.94    \\
 218&17 18 27.28& -22 57 39.7&     wise(171526-225432)     &  A(e)     & 10.72    & 10.28  \\
 219&17 24 24.51& -23 43 02.1&     SAO 185361              &    B9III  & 10.34    &10.00    \\
 220&17 15 12.80& -22 41 31.0&     CPD-22 11931            &    B      &  9.8     &      \\
 221&17 24 51.63& -23 55 33.9&     SAO 185369              &   B5III   &  9.07    & 8.77   \\
 222&17 26 22.22& -24 10 31.1&     SAO 185401              &   A3m     &  4.437   &4.166  \\
 223&17 09 10.80& -22 15 29.0&  wise(170911-221529)        & A(l)      & $>$14    &       \\
 224&17 12 23.44& -22 55 30.8&     SAO 185143              &  A5V      &  8.18    & 7.86   \\
 225&17 25 10.75& -24 36 18.7&     SAO 185375              &  B9       &   8.63   &8.39  \\
 226&17 18 59.09& -24 00 51.7&     SAO 185259              & A2III/IV  & 9.94     &9.75    \\
 227&17 20 42.59& -24 16 16.7&     SAO 185287              & B3V       &  9.03    &8.87  \\
 228&17 23 53.80& -24 51 25.5&    wise(172354-245126)      &    A(l)   & 11.6     & 10.9    
\end{tabular}
\end{table*}

\section{Results} 

The most evident result is that all the UV sources but one have been 
positively associated with a star of known or observed spectral type.
This very high percentage of identifications is the result of our
extensive  observational program.
For one source, N.207, no suitable candidate could be selected from
the photometry. For N.97, the bluest  candidate 
selected by the photometry was too faint to be observed with the FOSC 
camera.
 A globular cluster (M9) is 
identified as the optical counterpart of source N.102.  

Table 3 lists the distribution of the identified sources according to 
spectral type.
\begin{table}
\caption{Distribution of UV sources}
  \begin{tabular}{@{}rccccc@{}}
   %\tiny
$m_{UV}$ & B0 $-$ B9 & A0 $-$ A9 & F0 $-$ F9 & G0 $-$ G9 & No sp \\
5$-$6   & 2   & 1  & 0 & 0 & 0\\
6$-$7   & 2   & 3  & 0 & 0 & 0\\
7$-$8   & 10  & 3  & 0 & 1 & 0\\
8$-$9   & 10  & 7  & 0 & 0 & 0\\
9$-$10  & 23  & 17 & 1 & 0 & 0\\
10$-$11 & 18  & 41 & 1 & 3 & 0\\
11$-$12 & 9   & 46 & 4 & 1 & 2\\
12$-$13 & 5   & 13 & 1 & 0 & 1\\
13$-$14 & 0   & 2  & 0 & 0 & 0\\
14$-$15 & 0   & 0  & 0 & 0 & 0\\
$>$15  & 0   & 1  & 0 & 0 & 0\\ 
\hline
Total & 79   & 134& 7& 5 & 3

\end{tabular}
\end{table}
It is clear that almost all the stars belong to the B or A spectral types.
No star 
later than G was found. Only two sources, N.96 and N.152, are identified with 
an early G-type star. The [UV-v] color for both these stars is negative, 
while normal G stars are  much brighter in the visual range  than in the UV. 
Such a strong UV excess may be due to the presence of some coronal
activity or of a hot companion hidden by the late -type star.
 
Six UV sources are identified with Am  or Ap stars. These stars are known 
to have weak UV flux with respect to the regular A-type stars. While this
is true for the Am stars identified with the UV sources N.64 and N.222, 
the  [UV-v] colors of Ap  stars cover the range of normal A-type stars.     

Fig. 7 shows the distribution of the FAUST magnitude versus the instrumental
(UV--v) color. 
There are two points at the extremes of Fig. 7. 

The object at the extreme right, N.74, is the faintest UV
source detected in this sample and the error in its  UV flux is very large
($\simeq $ 45 \% ); this affects the error in the color index as well.
The point at the extreme left corresponds to source N.228. The (FAUST-v) 
color of this object is very negative, as expected for a very hot object 
or a  subdwarf. Below we will show that this object is indeed a 
white dwarf.
\begin{figure}
\hspace{10cm}
\centerline{\epsfxsize=3.5in\epsfbox{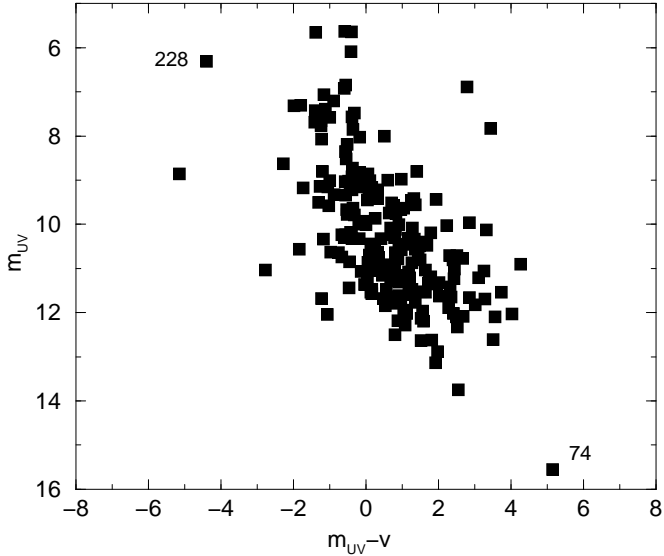}}
\caption{Distribution of the FAUST m$_{UV}$ magnitude versus the
color $m_{UV}-v$, for all the 227 identified sources.}
\end{figure}

\subsection{ A search for  Hidden Subluminous Stars}

No white dwarf (WD) or hot subdwarf (sd) stars were found as optical 
counterparts of the FAUST sources using the methods described above. 
 This result is similar to that found in the identification of 
UV sources in the Virgo FAUST field (Brosch et al. 1997). A deficiency in
hot evolved stars was found in other FAUST fields (e.g., Daniels et al. 2001).
This could be the result of the lack of such objects in the catalogues
from which possible counterparts are selected.

Hot subluminous stars are numerous among blue objects.
Systematic  colorimetric surveys, such as
the  Palomar Green (PG: Green et al. 1986) or the US Survey 
(Usher and Mitchell 1990),  show that these objects indeed dominate 
the population of blue stars down to B=16.5.

Furthermore, far-ultraviolet surveys such as those by the ROSAT Wide Field
Camera (Pye et al. 1995) and by the Extreme Ultraviolet Explorer (EUVE)
(Bowyer et al. 1994, 1996) 
discovered a substantial number (120) of hot WDs (Barstow                                                                 et al. 1994a,
Barstow et al. 1994b, Marsh et al. 1997, Burleigh et al. 1997).
The WDs detected by EUVE are very hot objects, with temperatures greater 
than several 10$^{4}$ K, whereas FAUST essentially covers lower effective
temperatures. 
At the limit magnitude of our sample of $m_{v} \simeq 12$, a WD would 
be a very luminous object ($M_{V} \sim$10). Very few such objects were
found by the PG survey.

   The number density of WD up to $M_{V} \sim$10 in the PG survey is 
about 0.04 per 1000 $pc^{-3}$ (Boyle 1989). With a maximal distance to 
detect WD with FAUST set at 25 pc, our sampled volume is 
$ \simeq $ 150 pc$^{3}$ with an expected total number of WDs to $M_{V} \sim$10
of 6 x 10$^{-3}$.
Furthermore, this prediction is an upper limit, since our sample of 
UV sources is not magnitude-limited.

In order to search for the missing sd/WD stars in our field we take 
advantage of the photometric and astrometric information provided by 
the Hipparcos satellite. We cross-identified our UV sources positions with 
objects in the Main Part of the Tycho catalog (ESA 1997). For all the UV stars 
under considerations a Tycho entry is available and a parallax value
is significant for 46 entries.
Table 4 gives the information retrieved from the 
Tycho catalog for all the stars and the derived distance (d) and absolute
magnitude (M${_V}$).
Table 4 lists  also the proper motion components in milliarsec/yr.

\begin{table*}
\caption{Hipparcos data}
  \begin{tabular}{@{}rrrrrrr@{}}
%\tiny
Id no.&$\pi$ (mas)$\pm$$\sigma$ & d(pc)& M${_V}$ & Sp type & 
$\mu$$_\alpha$(mas/yr) & $\mu$$_\delta$(mas/yr)\\
\\
 16 &  10.4$\pm$   6.4&   96.1&    2.92&   A3III    &    21.5$\pm$   8.1&  -11.6$\pm$   5.8 \\  
  18 &   8.3$\pm$   5.9&  120.4&    2.96&   F2V      &   -27.2$\pm$   5.8&  -15.8$\pm$   3.4 \\  
  20 & 148.5$\pm$  47.0&    6.7&   12.09&   A        &                   &                   \\  
  24 &  29.3$\pm$  22.8&   34.1&    7.66&   A0III    &   -39.9$\pm$  18.6&   15.4$\pm$  15.7 \\  
  25 &  56.9$\pm$  29.4&   17.5&    9.45&   B9III    &    10.2$\pm$  34.1&   33.2$\pm$  20.0 \\  
  32 &   9.3$\pm$   4.5&  107.5&    2.44&   F0IV     &   -19.0$\pm$   5.3&   -6.8$\pm$   2.6 \\  
  37 &  13.5$\pm$   8.8&   74.0&    4.45&   B9IV     &    21.5$\pm$  11.6&  -11.7$\pm$   8.0 \\  
  43 &  54.8$\pm$  35.0&   18.2&    9.19&   B9IV     &   -66.3$\pm$  32.8&  -95.6$\pm$  27.4 \\  
  58 &  65.8$\pm$  30.3&   15.1&    9.85&   B9III    &                    &                  \\  
  62 &  96.3$\pm$  59.1&   10.3&   10.95&   A(e)     &    -5.0$\pm$  46.1&  -76.0$\pm$  27.2 \\  
  65 & 118.8$\pm$  43.1&    8.4&   11.74&   A(l)     &    77.8$\pm$  35.6&  -17.6$\pm$  22.1 \\  
  66 &  90.9$\pm$  33.1&   11.0&   10.74&   A(e)     &                    &                  \\  
  68 &  15.7$\pm$  15.6&   63.6&    5.47&   Ap       &                    &                  \\  
  70 &   8.7$\pm$   5.4&  114.9&    2.55&   A7V      &                    &                  \\  
  74 & 117.4$\pm$  47.5&    8.5&   11.48&   A(e)     &   157.1$\pm$  49.2&   91.8$\pm$  28.6 \\  
  76 &  21.2$\pm$  15.3&   47.1&    6.43&   B8IV     &                    &                   \\  
  80 &  48.3$\pm$  18.9&   20.7&    8.49&   A7III/IV &   -34.5$\pm$  18.6&  -17.6$\pm$  12.3 \\  
  83 &  20.6$\pm$  19.4&   48.5&    5.83&   AV       &     6.5$\pm$  13.8&  -12.3$\pm$   9.4 \\  
  92 &  57.7$\pm$  51.3&   17.3&   10.35&   A(e)     &  -100.3$\pm$  64.3&    9.1$\pm$  43.8 \\  
  108&  17.1$\pm$  15.0&   58.4&    5.99&   B9V      &    28.5$\pm$  16.4&  -20.3$\pm$  12.0 \\  
  111&  67.1$\pm$  43.4&   14.9&   10.06&   A0IV     &   -53.2$\pm$  46.7&  -10.0$\pm$  38.9 \\  
  113&  71.3$\pm$  36.0&   14.0&   11.63&   A(l)     &    20.9$\pm$  34.7&  -42.0$\pm$  21.4 \\  
  123& 189.7$\pm$  43.4&    5.2&   12.64&   A(e)     &     5.1$\pm$  54.9& -103.0$\pm$  44.6 \\  
  126&  55.4$\pm$  27.5&   18.0&   10.07&   B9V      &    30.7$\pm$  29.8&  -22.2$\pm$  20.4 \\  
  127&  12.6$\pm$  11.4&   79.3&    4.88&   A0IV     &   -21.8$\pm$  12.2&  -14.1$\pm$   9.6 \\  
  139&  14.3$\pm$   6.8&   69.9&    4.09&   A3III    &     8.8$\pm$   8.8&  -40.1$\pm$   5.4 \\  
  148& 108.3$\pm$  61.3&    9.2&   11.85&   A(l), F  &   217.1$\pm$  49.5&   21.3$\pm$  42.7 \\  
  150& 100.3$\pm$  27.9&    9.9&   10.83&   A0V      &    13.7$\pm$  28.1&  -59.6$\pm$  21.2 \\  
  177&  44.0$\pm$  15.3&   22.7&    7.73&   A3V      &    30.1$\pm$  15.1&   -6.0$\pm$   9.7 \\  
     &  23.5$\pm$  18.5&   42.5&    6.72&   A8V      &    79.0$\pm$  18.9&  -38.1$\pm$  11.2 \\  
  178&  50.2$\pm$  24.6&   19.9&    9.09&   B9IV     &    -9.2$\pm$  31.1&  -28.8$\pm$  23.7 \\  
  180&  19.8$\pm$  11.5&   50.5&    7.38&   B9III    &   -24.4$\pm$  13.0&  -30.9$\pm$   8.8 \\  
  182&  40.8$\pm$  20.8&   24.5&    7.79&   A2III    &   -23.8$\pm$  17.4&   45.9$\pm$  14.3 \\  
  191&  36.1$\pm$  20.3&   27.7&    7.73&   A0V      &                    &                   \\  
  194&  49.9$\pm$  45.4&   20.0&    9.83&   A(e)     &   -88.4$\pm$  46.1&   14.4$\pm$  35.0 \\  
  198&  58.8$\pm$  20.1&   17.0&    9.18&   A2II/III &   -67.9$\pm$  22.7&  -11.0$\pm$  18.0 \\  
  202&  23.9$\pm$  16.0&   41.8&    6.52&   B9V      &                    &                   \\  
  205&  53.5$\pm$  49.2&   18.6&   10.03&   A(e)     &                    &                   \\  
  206& 114.0$\pm$  38.2&    8.7&   11.19&   A0III    &  -123.2$\pm$  47.2&  -35.7$\pm$  26.2 \\  
  210&   8.9$\pm$   4.3&  112.3&    2.22&   F2V      &    32.9$\pm$   5.4&   -2.7$\pm$   3.3 \\  
  211&  70.0$\pm$  47.3&   14.2&   10.60&   B(e)     &    25.0$\pm$  56.9& -107.1$\pm$  36.6 \\  
  213&  12.5$\pm$   5.9&   80.0&  $>$9.5&   A(l)     &   -27.6$\pm$   5.3&   -9.7$\pm$   2.9 \\  
  214& 224.4$\pm$  90.3&    4.4&   13.05&   A0IV     &                    &                   \\  
  220& 120.0$\pm$  57.2&    8.3&   12.13&  B         &  -92.4 $\pm$ 53.00& 102.90$\pm$ 41.50 \\  
  222&  36.8$\pm$   2.9&   27.1&    2.34&   A3m      &    -0.5$\pm$   3.2& -118.2$\pm$   2.0 \\  
  228& 139.1$\pm$  71.8&   7.18&   12.44&   A(l)     &   80.3 $\pm$ 59.80& -49.90$\pm$ 34.60 

\end{tabular}
\end {table*}
Fig. 8 shows the distribution 
of the proper motion modulus ($\mu$) with respect to the calculated distance (d). 
As expected for the low galactic latitude of the Ophiuchus field, the  sample 
is  composed  of stars with small proper motions, consistent with their
membership in the disk.
\begin{figure}
\hspace{10cm}
\centerline{\epsfxsize=3.5in\epsfbox{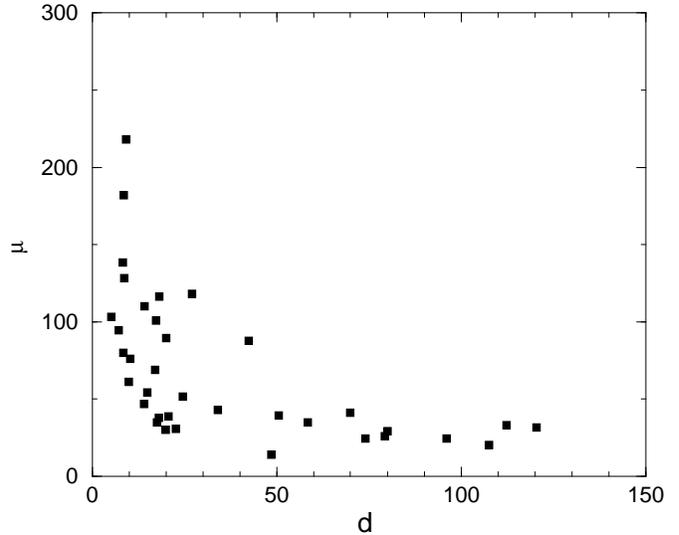}}
\caption{Distribution of the proper motion modulus $\mu$(mas) with
respect to the distance d(pc) for the sample
of stars retrieved from the Tycho catalog.}
\end{figure}

The parallax precision ($\sigma$$/$$\pi$) is very large for all the 
stars in Table 4 but one, namely SAO 1854091 identified with the source  N.222.
The median errors of the Hipparcos and Tycho parallaxes increase with apparent
magnitude. However, even at the  faintest Hipparcos  magnitudes (V $\ge$ 10.5), 
the median error amounts to few milliarsec(mas) (see The Hipparcos and 
Tycho catalogues, Fig 3.2.26 and Fig. 3.2.39).

For Tycho parallaxes  the median standard error is larger 
(see The Hipparcos and Tycho catalogues, Fig 3.3.14 and Fig. 3.3.16). 
For instance at Tycho magnitudes $V_{T}= 11$ it amounts to $\sim 50$
mas. The better precision for Hipparcos stars is implied by the different 
characteristic of the catalogues, where Hipparcos selected the brighter 
stars while the Tycho Catalog is a survey of all stars down to its
sensitivity limits. 

We neglected the presence of extinction in the line of sight and this 
omission impacts on the assigned absolute magnitude in Table 4. 
The interstellar extinction is very patchy (Schlegel et al. 1998)
and in the Ophiuchus region $A_{v}$ spans from 1 to more than 3 mag.
 
The parallax errors for almost all the entries in Table 4 are very large.
Observation errors 
that exceed a few mas are an indication of systematic problems in the
Hipparcos data reduction. 
A large error can indicate that the parallax is actually too small to
be measured, or can be due to the presence of an undetected multiple
system (Perryman et al. 1995)
\begin{figure}
\hspace{10cm}
\centerline{\epsfxsize=3.5in\epsfbox{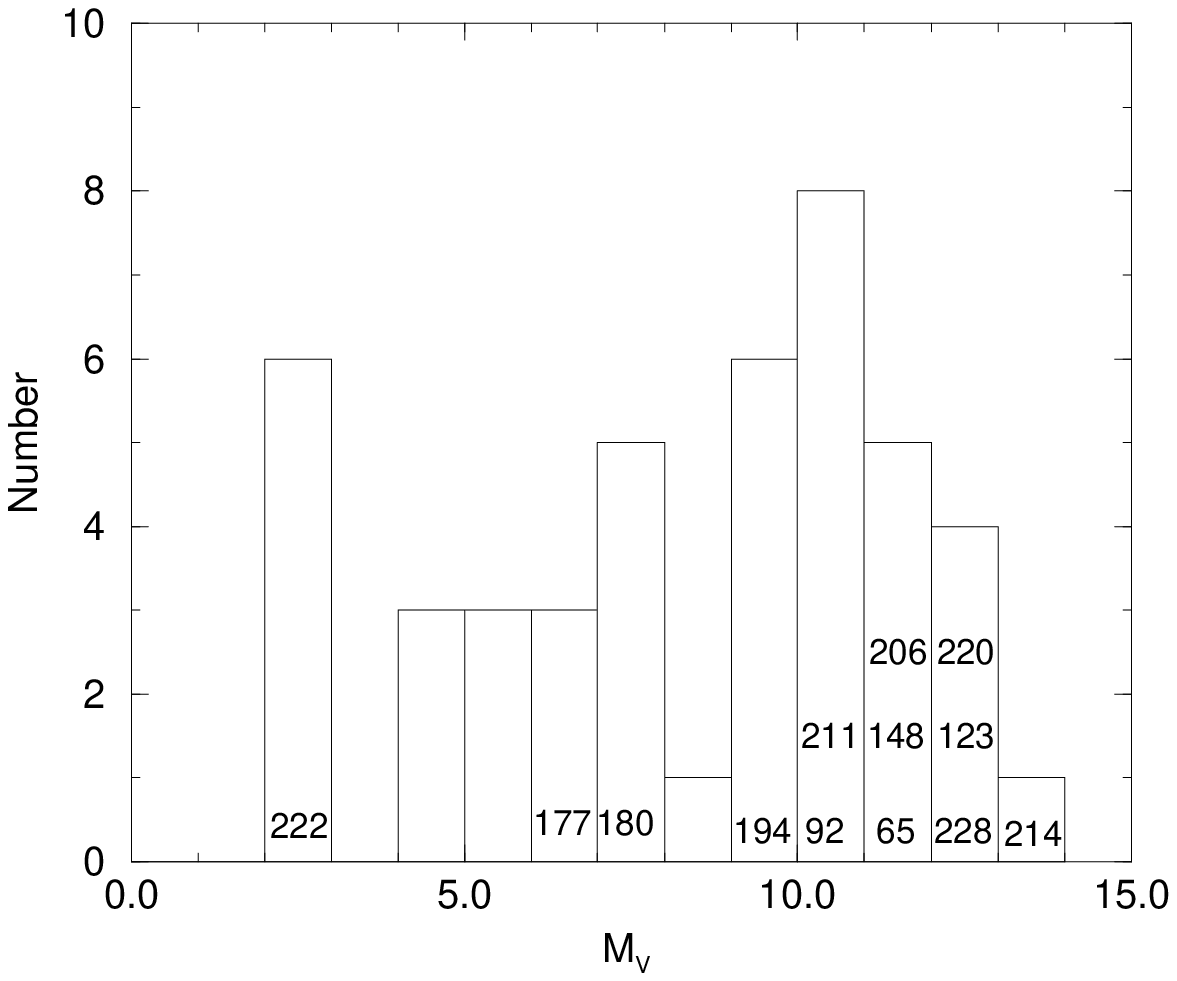}}
\caption{Histogram of the absolute magnitude for the UV sources
listed in the Tycho catalog.}
\end{figure}

Fig. 9 shows the histogram of the absolute magnitudes listed in Table 4.
The spectral classification of these stars spans types from B to early F,
and one would expect that M${_V}$ should be $\le$ 3.9 (Allen 1973) if these
would be main-sequence stars, while 
a significant number of objects in Fig. 9 have faint M${_V}$ values, 
as expected for
WD or subdwarf  stars. Note that the absolute V magnitude of a sdB subdwarf is 
$M_{V}=4.2 \pm 0.7$ (Heber 1986), while that of WDs ranges from 9.5 to 16.5.
Vauclair et al. (1997) analyzed the parallaxes obtained by the Hipparcos
mission with values obtained through ground observations, including 
the one-sigma measurements, and found a 2/3 fraction of agreements, 
compatible with the listed 1 sigma errors. 
Reid (1997) and Gratton et al (1997)  compared the parallax measurements 
from ground-based data with new ones obtained by the Hipparcos satellite
for a sample of subdwarfs. The Hipparcos parallaxes are systematically
smaller so that the derived absolute magnitude are on average brighter.
After applying the statistical correction for the Lutz-Kelker systematic 
bias towards overestimating parallax measurements, the mean difference in 
absolute magnitude is 0.53 $\pm$ 0.14 mag. 

The M${_V}$ calculated in Tab 4 for the stars of known spectral
type seems systematically underestimated with the respect 
to the value corresponding to their spectral type and
this effect is due to the inaccuracy of the parallax.
The astrophysical interest of Tycho parallaxes is restricted to
to the brighter stars (H\o g et al. 1997)
Hence, we may conclude that the search for subluminous stars 
on the basis of the parallax data of Table 4 is rather unreliable.

Many of the stars in Table 4 are catalogued as stars of known spectral type and 
some are stars observed at the Wise telescope during  the identification program.
It is possible, however, that some have not been recognized as subluminous stars
at  the low resolution and low signal-to-noise   quality of 
the spectra obtained for the FAUST sources  identification program.
 
In an effort to find the most probable subluminous stars among the  FAUST stars,
we calculated the predicted UV flux at the FAUST wavelength, performing 
synthetic photometry (section 5.2).
Furthermore, we embarked in a follow-up spectroscopic observational program 
of the candidates, i.e., of the objects showing a  FAUST magnitude much 
brighter than  the  predicted one (section 5.3).
The purpose of these observations was to cull the high gravity stars from the 
sample of ultraviolet-excess objects, by checking the Balmer line profiles
against models.

\subsection{Predicting the expected  UV magnitude}
In order to find the most probable subluminous stars hidden among
dwarf stars, we calculate the predicted UV flux of a star of a given spectral 
type and  luminosity, using synthetic photometry from spectral data.
A code has been developed (Formiggini, in preparation) that uses 
theoretical and  empirical stellar libraries (mainly from IUE).
For each specific wavelength and specific filter, the code is able to
predict the monochromatic magnitude 
\begin{equation}
UV_{\lambda}=-2.5 log f_{\lambda}+const
\end{equation}

If no spectral information is available for a star, the (B-V) color is used 
and the UV magnitude is predicted only on the basis of the photometric 
information. This code has  already been successfully 
tested for the Virgo FAUST sources listed in the Tycho catalog. 
The magnitudes predicted for the Virgo UV sources are well-correlated 
with the FAUST measured ones. A complete discussion of the results of this code, 
applied to all the FAUST  sources, will be presented in a forthcoming paper 
(Formiggini, in preparation). 
A preliminary version of this code has been tested 
on the photometric data set of the TD-1 catalog (Thompson et al. 1978) with
satisfactory results.

We applied the code to the sample of 46 stars identified as counterparts of the
Ophiuchus FAUST sources for which parallaxes are listed in the Tycho 
catalog. The visual and blue magnitudes listed in the Tycho catalog have been
used for the prediction of the UV magnitudes. The magnitude calculated
by the code can be used as an indication of the UV emission expected if the
object is  a main-sequence star. 
Fig. 10 shows the measured FAUST magnitudes with respect to the predicted ones.
The predicted magnitudes are fainter than the measured ones by one to two 
magnitudes; this systematic effect will be addressed in the forthcoming paper.   

\begin{figure}
%\hspace{10cm}
\centerline{\epsfxsize=3.5in\epsfbox{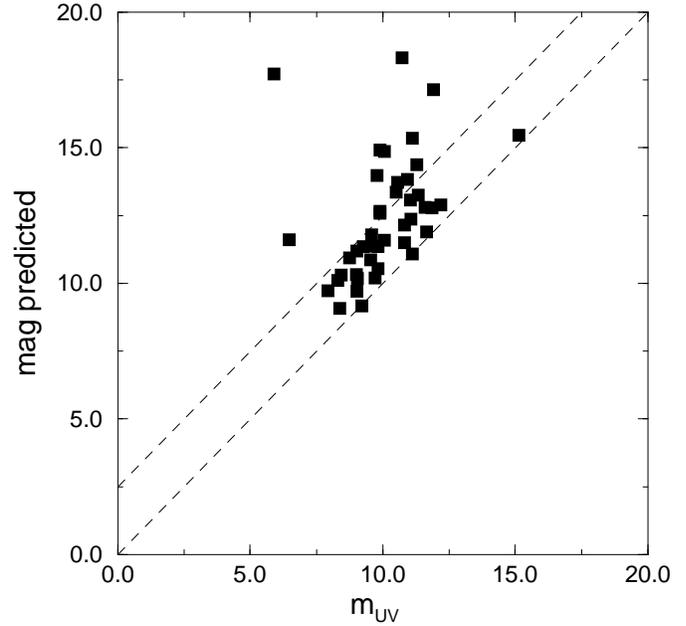}}
\caption{ FAUST magnidutes with respect to the predicted ones.}
\end{figure}
However, the relevant result of the prediction for the present discussion is 
the definition of a region inside the dotted lines in Fig. 10, which is 
populated by the  majority of the stars listed in Table 4.  There are 13 
stars outside this region for which FAUST measured a brighter UV 
magnitude than the one predicted on the basis of the photometric color. 

In Fig. 9 we marked, in each magnitude bin, the number of the UV source 
as listed in Table 1 of each  such star. 
According to their M${_V}$,  all these stars  but one, should belong to 
the subdwarf or white dwarf luminosity classes.
The source N. 222 is classified as an Am star and the  metallic lines affect
the (B--V) color used for the prediction. 
Actually, there are twelve ultraviolet excess sources with  M${_V}$ $\ge$ 5; 
these are the best candidates for hidden subluminous stars.

\subsection{ Spectroscopic Observations}
The twelve UV sources with M${_V}$ $\ge$ 5 and with a predicted FAUST magnitude
much brighter than the measured one, as explained in  the previous paragraph, were 
observed with the 1.0 meter telescope at the Wise Observatory.
The FOSC  was used with a  2" slit and a 600 g/mm grism. This yields a wavelength 
coverage from $\sim 3400 $\AA\ to $\sim 7100 $\AA\,
at a spectral resolution of $\sim 3.67$\AA\  per pixel. Although the spectral resolution
is not significantly higher than that used in the earlier identification
observations, the exposure times were longer in order to achieve a 
good signal-to-noise.

The spectra were reduced using standard techniques. Pixel-to-pixel sensitivity
variations were removed using flat-field spectra acquired at the beginning of 
each night, and  wavelength calibration was achieved using He-Ar arc spectra
observed  after each object spectrum. A bright star of known 
spectroscopical type was observed with the same  configuration as used for 
program stars. 
The spectra could not be flux calibrated, owing to the narrow slit compared to
the seeing size; some of the starlight was lost. Since the purpose of these 
observations is to cull high-gravity stars from the sample, only relative 
calibration in a restricted spectral segment is required for profile analysis.

\subsection{The line profiles   }
Table 5 lists the observed objects. The spectra of source N.92 were of bad 
quality and were discarded.
\begin{table*}
%\centering
\begin{minipage}{80mm}
\caption{ Candidates subluminous stars}
  \begin{tabular}{@{}rrrrrrl@{}}
%\tiny
Id no. & listed Sp. Type &m$_{v}$ & $\pi$ (mas)$\pm$$\sigma$ &UV-excess mag&
New classification \\
\\
65  &       A(l) & 10.7 & 118.8 $\pm$  43.2  & 5.0&sdB\\
92  &       A(e) &11.4  &  57.7 $\pm$  51.3  & 4.2&not observed\\
123 &      A(e)  &10.8  & 189.7 $\pm$  43.4  & 4.2&sdB\\
148 &     A(l),F &11.4  & 108.3 $\pm$  61.3  & 2.9&sdB\\
177 &       A8V  &9.8   &  23.5 $\pm$  18.5  & 2.7&not observed\\
180 &      B9III &10.9  &  19.8 $\pm$  11.5  & 4.8&not observed\\
194 &     A(e)   &10.4  & 49.9  $\pm$  45.4 & 7.6&sdB\\
206 &      A0III &10.5  & 114.  $\pm$  38.2 & 2.9&sdB\\
211 &      B(e)  &11.   & 70.0  $\pm$  47.3  & 3.1&sdB\\
214 &      A0IV  &10.9  & 224.4 $\pm$  90.3 & 3.2&wd\\
220 &      B     &10    & 120.  $\pm$  57.2 &5.2& B\\
228 &      A(l)  &10.9  & 139.1 $\pm$  71.8  &11.8&wd
\end{tabular}
\end{minipage}
\end{table*}
A visual inspection  of the higher quality spectra  reveals the presence of 
moderately broadened  Balmer lines for all the objects observed but one, N. 220.
Broad and shallow Balmer lines are the characteristic allowing us
to isolate the subdwarf class among the hot stars, since line profiles depend
very sensitively on the value of gravity (g) and on temperature (T$_{eff}$).
The absence of the absorption line  He II 4686 indicates that the sdO
class cannot be assigned to one of the nine objects discussed here.

A reliable determination of stellar atmospheric parameters can  be obtained by
simultaneously fitting observed line profiles to appropriate model atmospheres. 
This has been done by many authors using grids of model stellar atmospheres.
For instance, such models, in local thermodynamic equilibrium(LTE) and non-LTE,
and taking into account the  blanketing process, 
have been computed and used by Saffer et al. (1994) for sdB stars. However the 
models used by  Saffer et al. have not been published.

We found in the literature available models which  are the results of a grid of LTE  and 
of some NLTE and blanketed pure-hydrogen models, calculated by Wesemael et al. (1980). 
These  pioneering models are still valid for the aim of our analysis.
The influence of NLTE effects and metal-line blanketing on hydrogen line 
profiles of high gravity stars has been investigated by Napiwotzki 
(1997) and by Barstow et al. (1998). The NLTE effects are significant only 
at highest temperatures and at low log g, and the differences are detected only
in the core and  only at high spectral resolution (Saffer 1994).
The metal-line blanketing is more dominant at the the temperature
of sdB (see Napiwotzki (1997), fig 10) and should be taken into account 
for an accurate  determination of the atmosphere parameters.
More recently, a completely new grid of stellar atmosphere models have 
been calculated (Barstow et al 2001) using the code TLUSTY (Hubeny and Lanz 1995).

Since the aim of this work is to isolate the subluminous stars, without 
embarking in an accurate determination of the atmospheric parameters, we 
compared the Balmer profiles of the observed spectra with the LTE 
unblanketed model profiles published by Wesemael et al. (1980). 

The theoretical line profiles were convolved with the instrumental resolution 
profile. This profile was determined from the emission lines produced by the 
He-Ar comparison lamp, and was approximated by a Gaussian profile of 
FWHM $\sim 6.2  $\AA\ . 
No simulated continuum, such as a Poisson noise, was added to the model 
profiles, and no convolution with a rotation broadening function was 
performed. Rotation could affect the line cores, while we are interested in 
the change in the line wings due to effective  temperature and surface 
gravity variations. Saffer et al (1994) found that only for very large 
velocities ($\ge 100$ \kms) the stellar rotation can perturb the line wings.

Around the Balmer line rest wavelength, two symmetrical $100$\AA\ segments 
of the observed spectrum were extracted, sufficiently far from the line 
to define adequately the continuum on either side of it.
The local slope of the continuum was subtracted from the spectrum.  
The observed and theoretical line profiles were interpolated and rectified 
to a linear continuum in a consistent manner. 
\begin{figure*}
\begin{minipage}{80mm} 
%\hspace{10cm}
\centerline{\epsfxsize=4.0in\epsfbox{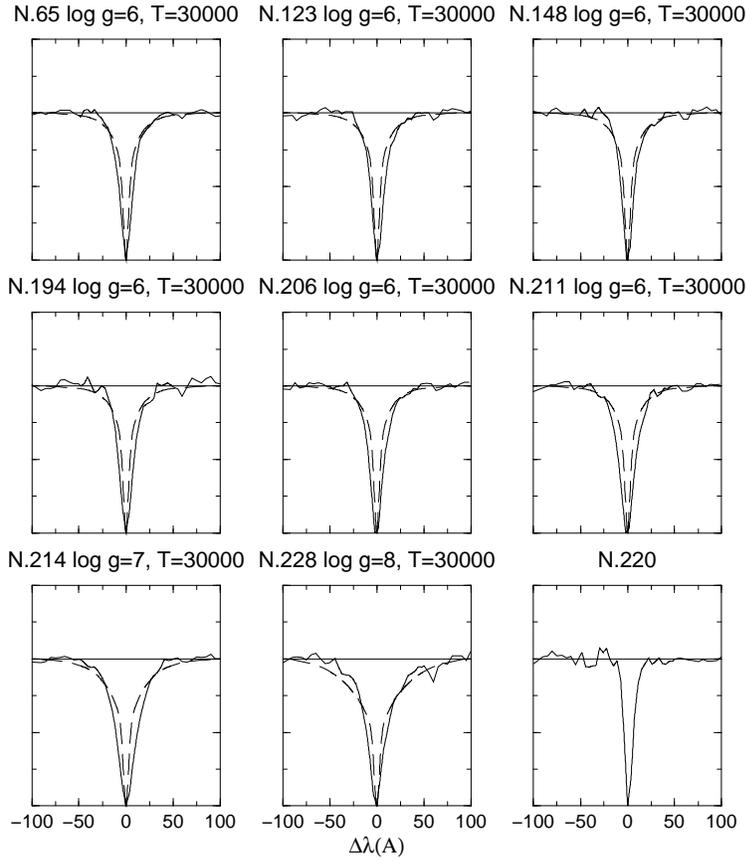}}
\caption{ The H$\beta $  profile observed (line) and the
model atmospheres profile (dashed) at the marked temperature and gravity,
for the nine candidates.}
\end{minipage}
\end{figure*}

Fig 11 shows the H$\beta $ line profile for the nine objects observed. 
In some profiles there is a residual systematic deviation in the continuum
spectral slope between the model and the data, 
which should be dealt with  for an accurate analysis.
The dashed line in Fig 11 is  the model profile from Wesemael et al. (1980) for 
log g=6 and $T_{eff}= 30000 $K. For N.214 and N.228 the dashed line 
corresponds to a  higher gravity model, while the source N.220 shows a 
narrow line profile. 

A proper model atmosphere analysis requires a simultaneous line-by-line analysis. 
The  models in Fig 11 
are only indicative and we  do not claim that this fitting is a final and authoritative 
determination of the exact temperature or/and gravity. We can only state that 
low-gravity models, such as for dwarf stars, are inappropriate for all the 
candidates observed but N.220.

Another approach is to compare the spectral profile with real stellar atmospheres,
such as the spectrum of a dwarf star and of a white dwarf observed under
the same conditions as  the candidate stars. 
We observed WD 2032+248 (G186-31), classified as  a DA 
white dwarf with atmospheric parameters $T_{eff}= 19980 $K and log g=7.83
(Bergeron et al. 1992) , and the A0 V  star BD +26 2606, which was observed every 
night. Fig 12a and Fig 12b  show the  H$\beta $  line profile  of BD +26 2606 
superposed  to that for the objects N.211 and N.228. In Fig 12c the H$\beta $  
line profile for  WD 2032+248 is compared with that of N.228. 
These comparisons enhance the reliability of the classification of the candidates
as sdB/WD stars.

\begin{figure*}
\begin{minipage}{80mm}
%\hspace{10cm}
\centerline{\epsfxsize=4.0in\epsfbox{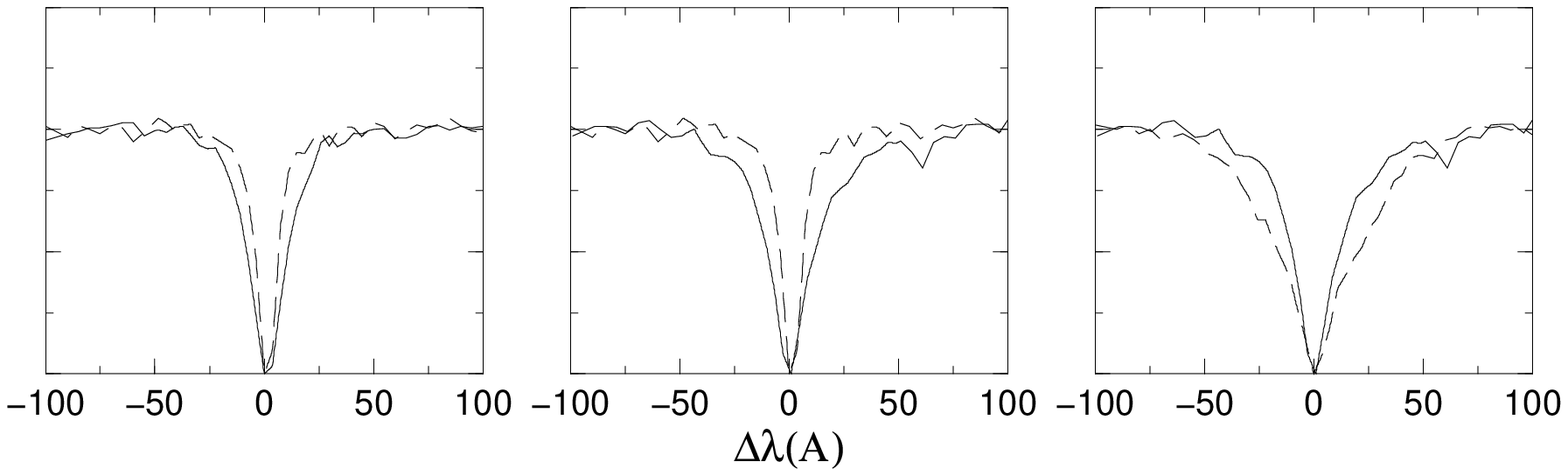}}
\caption{The  H$\beta $  profile or BD+26 2606 (dashed) compared a) with that 
for N.211 (line)  and b) with that for N.228(line): c) The H$\beta $  profile 
for WD 2032+248 compared with that for N.228 (line).}
\end{minipage}
\end{figure*}

Our spectral data are a fair confirmation of the presence of high-gravity 
atmospheres, i.e., of the presence of a substantial number of subluminous stars 
in the Ophiuchus FAUST sample. Two of the observed candidates can be fitted
with atmospheres of log g $\ge 7 $, as for white dwarfs stars, and
six are sdB stars.

Many sdB stars are members of binary systems with cool companions. 
Analyzing the spectral distribution of a  sample of 34 sdB, Aznar Cuadrado
and Jeffrey (2001) estimate the sdB binary fraction  between 56 \% to
100\% that have a main sequence star companion. The presence of an undetectable
binary companion could indeed be the source of the inaccuracy of the astrometric data.

The spatial distribution of sdB is important for a description of their
evolutionary scenario and for the links with other types of stars. sdB stars
are one of the evolutionary channels and direct progenitors to white dwarfs.
The space density of sdB at bright magnitudes is poorly defined. Complete samples
are needed for such investigations, or alternatively a reliable estimate of the
completeness of such samples. However the uncertainty on the scale-height 
distribution adopted, uniform or exponential, make the completeness correction 
quite difficult.

The PG survey is one of the largest and most studied survey of sdBs.
It is, however,  generally accepted that the PG sample is quite incomplete at
bright magnitudes ($\le$ B=12.5). Villeneuve et al (1995) 
review the problem of completeness  and give a revised space density of
  3 $\pm$ 1 *10$^{-7}$ pc$^{-3}$ and  an exponential scale height of the 
order of 450 $\pm$ 150 pc,  or alternatively an isothermal disk scale height
of  600 $\pm$ 150 pc.
Mitchell (1998) claims a mix of disk and halo populations. A significant 
halo component for sdB stars supports the hypothesis of similar spatial 
distributions for sdB and sdO stars and hence the possibility of an 
evolutionary process relating sdB and sdO.

A comprehensive search for previously unrecognized subdwarfs in the 
Hipparcos catalogue has been carried out by Reid at al (2001).
Among the stars with parallaxes measured to a precision better than 
15 per cent, only few (nine) such stars were discovered. 
Our analysis  shows that combining the UV information with 
even  poorly determined parallaxes, some new sdB are detected, 
and that indeed more such bright stars are still unrecognized in the 
existing samples. We are extending this analysis to the other
FAUST fields previously identified at the Wise Observatory.
  
\subsection{Possible members of the Ophiuchus  molecular cloud}

The Hipparcos distance of four stars in Table 4 is comparable to the distance 
of the Ophiuchus molecular cloud with which they appear associated, and 
their position is  about 12$^o$  away from the $\rho$ Ophiuchus core. Their
mean transverse velocity, calculated from the proper motion, is 14 \kms.

Star formation in  molecular clouds  occurs in two different environments;
in small, isolated and dense regions distributed throughout the molecular cloud, such
as in Orion, or  in a single massive concentration of dense gas, such as the
$\rho$ Ophiuchi core (Lada et al. 1993). The accumulation of dense  gas evolves
through star formation, to form a embedded cluster of stars. Dynamical relaxation process,
such as  energy equipartition, leads to the massive stars releasing potential energy
and sinking towards the center while the less massive stars gain kinetic energy
and "evaporate" leaving the cluster and the cloud.

The four stars which have the distance of the Ophiuchus cloud have spectral 
type later than A7 and are not massive. We suggest that these four stars could 
have escaped the young cluster by this process of  stellar-dynamical evaporation 
and migrated far away from the region of their formation.
At a distance of 125 pc, and with a mean velocity of 14 \kms, the 
evaporation process took place some 3 Myr ago.

\section{Conclusions}

We have analyzed the UV FAUST image in the direction of the
Ophiuchus molecular cloud and detected 228 UV sources.
Optical identification was presented for all sources.
Multicolor broad-band photometry was used in
selecting the candidates, followed by low-resolution spectroscopy. 
Spectral types are presented for 60 stars newly identified as optical
counterparts of the UV sources. Most of the UV sources  were 
identified as early-type stars. 
The Hipparcos/Tycho information was used to calculate the absolute
magnitude of these stars. Some 39 UV sources are associated with
hot sub-luminous stars, however, the large errors in the parallax 
measurements prevent a conclusive classifications of them as sd/WD stars.

Synthetic photometry of spectral data was performed in order to predict 
the expected UV emission, on the basis of the photometric information
from Tycho. The comparison of the predicted emission with the FAUST 
measured magnitudes allowed us to select twelve stars as  highly 
probable hot stars.
High signal-to-noise spectra were obtained for nine of these stars and 
Balmer line profiles were compared with the prediction of atmosphere 
models and with the spectrum of  real stellar atmospheres.
Among the nine candidates,  six were classified as previously 
unrecognized sdB stars and two as  white dwarfs.
The lack of completeness of our sample prevented us from 
estimating of the local density of subdwarfs. However, our result indicates 
that more bright subluminous stars are missed in the existing samples.

\section*{Acknowledgments}

The UV astronomy effort at Tel Aviv is supported  by special grants to develop a 
space UV astronomy experiment  (TAUVEX) from the Ministry of Science, and from
the Austrian Friends of Tel Aviv University. Part of this study was supported
by a grant from the US-Israel Bi-national Science Foundation. 
Observations at the Wise Observatory
are supported by a Center for Excellence Grant from the Israel Science Foundation. 
This paper made use of the SIMBAD data base operated at CDS, Strasbourg, France. 
We acknowledge help with this project, at various stages, from Ms. Susanna Steindling,
Ms. Orly Kovo and Mr. E. Goldberg and discussions with Martin Barstow.

\label{lastpage}

\end{document}